\documentclass[11p]{article} 
\usepackage{amsmath}
\usepackage{amsfonts}
\usepackage{amssymb}
\usepackage{lmodern}

\usepackage[hyphens]{url}

\usepackage{times}

\usepackage[pdftex]{graphicx}
\usepackage{epstopdf}

\usepackage{color}

\usepackage{multirow}

\usepackage{changepage}

\usepackage[authoryear]{natbib} 

\usepackage{array}

\usepackage{rotating}

\usepackage{bbm}

\usepackage{latexsym}

\usepackage[lofdepth,lotdepth]{subfig}

\usepackage{arydshln}

\DeclareGraphicsExtensions{.eps}

\usepackage{appendix}

\usepackage{caption}

\usepackage{lineno}
\usepackage{setspace}

\newcommand{\captionfonts}{\it\normalsize}

\makeatletter  

\long\def\@makecaption#1#2{
	
	\vskip\abovecaptionskip
	
	\sbox\@tempboxa{{\captionfonts #1: #2}}%
	
	\ifdim \wd\@tempboxa >\hsize
	
	{\captionfonts #1: #2\par}
	
	\else
	
	\hbox to\hsize{\hfil\box\@tempboxa\hfil}%
	
	\fi
	
	\vskip\belowcaptionskip}

\makeatother

\newlength\savedwidth

\newcommand\thickhline{\noalign{\global\savedwidth\arrayrulewidth\global\arrayrulewidth 2pt}
	
	\hline
	
	\noalign{\global\arrayrulewidth\savedwidth}}

\RequirePackage[normalem]{ulem}
\RequirePackage{color}\definecolor{RED}{rgb}{1,0,0}\definecolor{BLUE}{rgb}{0,0,1}

\providecommand{\DIFaddbegin}{} 
\providecommand{\DIFaddend}{} 
\providecommand{\DIFdelbegin}{}
\providecommand{\DIFdelend}{}

\providecommand{\DIFaddbeginFL}{}
\providecommand{\DIFaddendFL}{} 
\providecommand{\DIFdelbeginFL}{} 
\providecommand{\DIFdelendFL}{}
\newcommand{\DIFscaledelfig}{0.5}

\RequirePackage{settobox} 
\RequirePackage{letltxmacro}
\newsavebox{\DIFdelgraphicsbox} 
\newlength{\DIFdelgraphicswidth} 
\newlength{\DIFdelgraphicsheight} 

\LetLtxMacro{\DIFOincludegraphics}{\includegraphics} 
\newcommand{\DIFaddincludegraphics}[2][]{{\color{blue}\fbox{\DIFOincludegraphics[#1]{#2}}}} 
\newcommand{\DIFdelincludegraphics}[2][]{
	\sbox{\DIFdelgraphicsbox}{\DIFOincludegraphics[#1]{#2}}
	\settoboxwidth{\DIFdelgraphicswidth}{\DIFdelgraphicsbox} 
	\settoboxtotalheight{\DIFdelgraphicsheight}{\DIFdelgraphicsbox}
	\scalebox{\DIFscaledelfig}{
		\parbox[b]{\DIFdelgraphicswidth}{\usebox{\DIFdelgraphicsbox}\\[-\baselineskip] \rule{\DIFdelgraphicswidth}{0em}}\llap{\resizebox{\DIFdelgraphicswidth}{\DIFdelgraphicsheight}{
				\setlength{\unitlength}{\DIFdelgraphicswidth}
				\begin{picture}(1,1)
				\thicklines\linethickness{2pt}
				{\color[rgb]{1,0,0}\put(0,0){\framebox(1,1){}}}
				{\color[rgb]{1,0,0}\put(0,0){\line( 1,1){1}}}
				{\color[rgb]{1,0,0}\put(0,1){\line(1,-1){1}}}
				\end{picture}
			}\hspace*{3pt}}} 
}
\LetLtxMacro{\DIFOaddbegin}{\DIFaddbegin}
\LetLtxMacro{\DIFOaddend}{\DIFaddend} 
\LetLtxMacro{\DIFOdelbegin}{\DIFdelbegin} 
\LetLtxMacro{\DIFOdelend}{\DIFdelend} 
\DeclareRobustCommand{\DIFaddbegin}{\DIFOaddbegin \let\includegraphics\DIFaddincludegraphics} 
\DeclareRobustCommand{\DIFaddend}{\DIFOaddend \let\includegraphics\DIFOincludegraphics} 
\DeclareRobustCommand{\DIFdelbegin}{\DIFOdelbegin \let\includegraphics\DIFdelincludegraphics} 
\DeclareRobustCommand{\DIFdelend}{\DIFOaddend \let\includegraphics\DIFOincludegraphics}
\LetLtxMacro{\DIFOaddbeginFL}{\DIFaddbeginFL} 
\LetLtxMacro{\DIFOaddendFL}{\DIFaddendFL}
\LetLtxMacro{\DIFOdelbeginFL}{\DIFdelbeginFL} 
\LetLtxMacro{\DIFOdelendFL}{\DIFdelendFL} 
\DeclareRobustCommand{\DIFaddbeginFL}{\DIFOaddbeginFL \let\includegraphics\DIFaddincludegraphics} 
\DeclareRobustCommand{\DIFaddendFL}{\DIFOaddendFL \let\includegraphics\DIFOincludegraphics}
\DeclareRobustCommand{\DIFdelbeginFL}{\DIFOdelbeginFL \let\includegraphics\DIFdelincludegraphics}
\DeclareRobustCommand{\DIFdelendFL}{\DIFOaddendFL \let\includegraphics\DIFOincludegraphics}

\usepackage[utf8]{inputenc}

\usepackage[T1]{fontenc}

\begin{document}
	
	\thispagestyle{empty}

	\markboth{ }{ }

	\ \vspace{-0mm}\\
	\singlespacing
	{\LARGE Perceptual decision making: Biases in post-error reaction times explained by attractor network dynamics}
	\ \\~\\
	{\bf \large Kevin Berlemont$^{\displaystyle 1}$, and Jean-Pierre Nadal$^{\displaystyle 1, \displaystyle 2}$}\\~\\
	{$^{\displaystyle 1}$ Laboratoire de Physique Statistique, \'Ecole Normale Sup\'erieure, PSL University, Universit\'e Paris Diderot, Universit\'e Sorbonne Paris Cit\'e, Sorbonne Universit\'e, CNRS, 75005 Paris, France.}\\
	{$^{\displaystyle 2}$ Centre d'Analyse et de Math\'ematique Sociales, \'Ecole des Hautes \'Etudes en Sciences Sociales, PSL University, CNRS, 75006 Paris, France.}\\

	{\bf Abstract } 
	Perceptual decision-making is the subject of many experimental and theoretical studies.
    Most modeling analyses are based on statistical processes of accumulation of evidence. In contrast, very few works confront attractor network models' predictions with empirical data from continuous sequences of trials.
	Recently however, numerical simulations of a biophysical competitive attractor network model have shown that such network can describe sequences of decision trials and reproduce repetition biases observed in perceptual decision experiments. Here we get more insights into such effects by considering an extension of the reduced attractor network model of Wong and Wang (2006), taking into account an inhibitory current delivered to the network once a decision has been made. We make explicit the conditions on this inhibitory input for which the network can perform a succession of trials, without being either trapped in the first reached attractor, or losing all memory of the past dynamics. 
    We study in details how, during a sequence of decision trials, reaction times and performance depend on the nonlinear dynamics of the network, and we confront the model behavior with empirical findings on sequential effects. 
	Here we show that, quite remarkably, the network exhibits, qualitatively and with the correct orders of magnitude, post-error slowing and post-error improvement in accuracy, two subtle effects reported  in behavioral experiments
   {\it in the absence of any feedback} about the correctness of the decision. 
    Our work thus provides evidence that such effects result from  intrinsic properties of the nonlinear neural dynamics.

		\subsection*{Significance statement}
	
		Much experimental and theoretical work is being devoted to the understanding of the neural correlates of perceptual decision making.  In a typical behavioral experiment, animals or humans perform a continuous series of binary discrimination tasks. To model such experiments, we consider a biophysical decision-making attractor neural network, taking into account an inhibitory current delivered to the network once a decision is made.
		Here we provide evidence that the same intrinsic properties of the nonlinear network dynamics underpins various sequential effects reported in experiments. Quite remarkably, in the absence of feedback on the correctness of the decisions, the network exhibits post-error slowing (longer reaction times after error trials) and post-error improvement in accuracy (smaller error rates after error trials).
			
		\section*{Introduction}
	
		Typical experiments on perceptual decision-making consist of series of successive trials separated by a short time interval, in which performance in identification and reaction times are measured. The most studied protocol is the one of Two-Alternative Forced-Choice (TAFC) Task -- see e.g.	\cite{Ratcliff1978,Laming1979,Vickers1979,Townsend1983,Busemayer1993,Shadlen1996,Usher2001,Ratcliff2004}. 
		Several studies have demonstrated strong serial dependence in perceptual decisions between temporally close stimuli~\citep{Fecteau2003,Jentzsch2009,Danielmeier2011}. Such effects have been studied in the framework of statistical models of accumulation of evidence~\citep{Dutilh2011}, the most common theoretical approach to perceptual decision-making, see e.g. \cite{Ratcliff1978,Ashby1983,Shadlen2006,Ratcliff2008,Bogacz2006},
		 or with a more complex attractor network with additional memory units  specifically implementing a biasing mechanism~\citep{Gao2009}.
		
		\cite{wang2002probabilistic} proposed an alternative approach to the modeling of perceptual decision making based on a biophysical cortical network model of leaky integrate-and-fire neurons. The model is shown to account for random dot experiments results of~\cite{Shadlen2001,Roitman2002}. This decision-making attractor network has been first studied in the context of a task requiring to keep in memory the last decision. This working memory effect is precisely achieved by having the network activity trapped into an attractor state. However, in the context of consecutive trials, the neural activities have to be reset in a low activity state before the onset of the next stimulus. \cite{Bonaiuto2016} have considered a parameter range of weaker excitation where the working memory phase cannot be maintained.
		The main result is that the performance of the network is biased towards the previous decision, an effect which decreases with the inter-trial time. Due to the slow relaxation dynamics in the model, the authors only study inter-trial times longer than $1.5$~s. However, sequential effects have been reported for shorter inter-trial times, such as $500$~ms in \cite{Laming1979,Danielmeier2011}.   		
		Instead of decreasing the recurrent excitation, an alternative is to introduce an additional inhibitory input following a decision~\citep{Lo2006,Engel2015,Bliss2017a}.
		\cite{Lo2006} have proposed such a mechanism to account for the control of the decision threshold.

			The purpose of the present paper is to revisit this issue of dealing with sequences of successive trials within the framework of attractor networks    with a focus on inter-trial times as short as $500$~ms. 
			We do so by taking advantage of the reduced model of \cite{Wang:2006} which is amenable to mathematical analysis. This model consists of a network of two units, representing the pool activities of two populations of cells, each one being specific to one of the two stimulus categories. \cite{Wang:2006} derive the equations of the reduced model and choose the parameters values in order to preserve as much as possible the dynamical and behavioral properties of the original model.
			In line with \cite{Lo2006}, we take into account an inhibitory current originating from the basal ganglia, occurring once a decision has been made.  
            We explore  how the network nonlinear dynamics leads to  serial dependence effects in TAFC tasks, 
           and  compare with empirical findings 
            such as sequential bias in decisions ~\citep{Cho2002} or post-error adjustments~\citep{Danielmeier_etal2011,Danielmeier2011}. 		Our main finding is that
            the model reproduces two main post-error adjustments observed in the absence of feedback on the correctness of the decision: post-error slowing (PES) and post-error improvement in accuracy (PIA), with PES consisting of longer reaction times, and PIA of smaller error rates, for trials following a trial with incorrect decision.
			We thus provide evidence that such effects result from    nonlinearities in the neural dynamics.
             
			\section*{Materials and Methods}
		
			We are interested in modeling experiments where a subject has to decide whether a stimulus belongs to one or the other of two categories, hereafter denoted $L$ and $R$. A particular example 
            is the one of random dot experiments~\citep{Shadlen2001,Roitman2002}, where a monkey performs a motion discrimination task in which it has to decide whether a motion direction, embedded into a random dot motion, is towards left ($L$) or right ($R$). 
            The general case is the one of categorical perception experiments, in which one can control the degree of ambiguity of the stimuli -- e.g. psycholinguistics experiments with stimuli interpolating between two phonemes~\citep{Liberman1957}, visual categorization experiments with continuous morphs from cats to dogs~\citep{Freedman2003}, etc.   
              We focus on Two Alternative Forced-Choice (TAFC) protocols in which no feedback is given on the correctness of the decisions.
            
			We consider a decision-making recurrent network of spiking neurons governed by local excitation and feedback inhibition, as introduced and studied in~\cite{Compte2000} and~\cite{wang2002probabilistic}. 
			Since mathematical analysis is harder to be performed for such complex networks, without a high level of abstraction ~\citep{Miller2013}, 
			one must rely on simulations which, themselves, can be computationally heavy. For our analysis, we make use of the reduced firing-rate model of~\cite{Wang:2006} obtained by a systematic reduction of the detailed biophysical attractor network model.  The reduction aimed at faithfully reproducing not only the behavioral behavior of the full model, but also neural firing rate dynamics
            and the output synaptic gating variables. This is done within a mean-field approach, with calibrated simplified F-I curves for the neural units,  
				and in the limit of slow NMDA gating variables motivated by neurophysiological data. The full details can be found in ~\cite{Wang:2006} (main text and Supplementary Information).

			Since this model has been built to reproduce as faithfully as possible the neural activity of the full spiking neural network, it can be used as a proxy for simulating the full spiking network~\citep{Engel2011,Deco2013,Engel2015}. Here, we mainly make use of this model to gain better insights into the understanding of the model behavior.  In particular, one can conveniently represent the network dynamics in a 2-d phase plane and rigorously analyze the network dynamics~\citep{Wang:2006}.

			\subsection*{A reduced recurrent network model for decision-making.}
			\label{wongwang}
\begin{figure}[!ht]
				\centering
				 \includegraphics[width=1.0\textwidth]{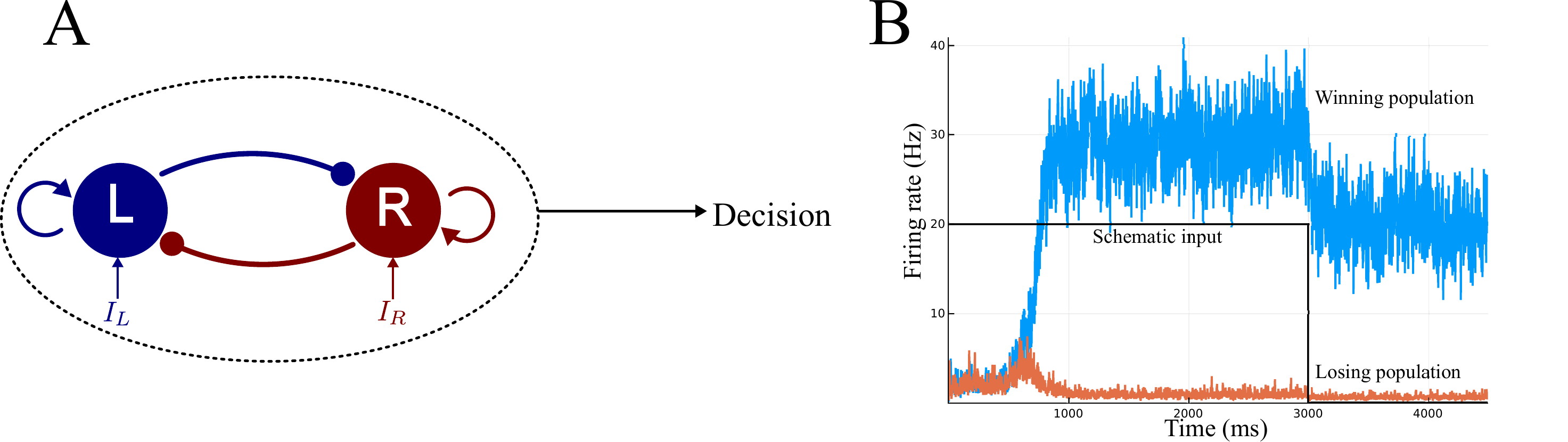} 
				\caption{{\bf Two-variable model of~\cite{Wang:2006}.}			(A) Reduced two-variable model~\cite{Wang:2006} constituted of two neural units, endowed with self-excitation and effective mutual inhibition. (B) Time course  of the two neural activities during a decision-making task.  
							At the beginning the two firing rates are indistinguishable. 
							The firing rate that ramps upward (blue) represents the winning population, the orange one the losing population. A decision is made when one of the firing rate crosses the threshold of $20$ Hz. 
							The black line represents the duration of the selective input corresponding to the duration of accumulation of evidence until the decision threshold is reached. This model shows working memory through the persistent activity in the network after the decision is made. 
						}
				\label{fig1}
			\end{figure}

            We first present the architecture without the corollary discharge (\cite{Wang:2006}, Fig. \ref{fig1} panel A), which consists of two competing units, each one representing an excitatory neuronal pool, selective to one of the two categories, $L$ or $R$. The two units inhibit one another, while they are subject to self-excitation. The dynamics is described by a set of coupled equations for the synaptic activities $S_L$ and $S_R$ of the two units $L$ and $R$:
			\begin{equation}
			i \in \{L,R\}, \;\;\dfrac{\mathrm{d} S_i}{\mathrm{d} t} = - \dfrac{S_i}{\tau_S} + \left( 1-S_i \right) \gamma f \left( I_{i,tot} \right)
			\label{dSidt}
			\end{equation}
			The synaptic drive $S_i$ for pool $i \in \{L,R\}$ corresponds to the fraction of activated NMDA conductance, and $I_{i,tot}$ is the total synaptic input current to unit $i$. The function $f$ is the effective single-cell input-output relation~\citep{Abbott2005a}, giving the firing rate as a function of the input current:
			\begin{equation}
			f \left( I_{i,tot} \right) = \dfrac{a I_{i,tot} - b}{1 - \exp \left[ -d \left( a I_{i,tot} - b\right)\right]}
			\label{fI}
			\end{equation}
			where $a, b, d$ are parameters whose values are obtained through numerical fit.
			
			The total synaptic input currents, taking into account the inhibition between populations, the self-excitation, the background current and the stimulus-selective current can be written as:
			\begin{align}
			I_{L,tot} &= J_{L,L} S_L - J_{L,R} S_R + I_{stim,L} + I_{noise,L} \\
			I_{R,tot} &= J_{R,R} S_R - J_{R,L} S_L + I_{stim,R} + I_{noise,R}
            \label{ILRtot}
			\end{align}
			with $J_{i,j}$ the synaptic couplings ($i$ and $j$ being $L$ or $R$). The minus signs in the equations make explicit the fact that the inter-units connections are inhibitory (the synaptic parameters $J_{i,j}$ being thus positive or null). The term $I_{stim,i}$ is the stimulus-selective external input. If $\mu_0$ denotes the strength of the signal, the form of this stimulus-selective current is: 
			\begin{eqnarray}
				I_{stim,L}= J_{A,ext} \mu_0 \left( 1 \pm c\right) \nonumber \\
I_{stim,R}= J_{A,ext} \mu_0 \left( 1 \mp c\right)
				\label{Istim}
				\end{eqnarray}
				The sign, $\pm$, is positive when the stimulus favors population $L$, negative in the other case. The quantity $c$,  between $0$ and $1$, 
				gives the strength of the signal bias. It quantifies the 
				coherence level of the stimulus. For example,  in the random dot motion framework, it corresponds to the fraction of dots contributing to the coherent motion.  In the following, we will give this coherence level in percent. 
			Following~\cite{wang2002probabilistic}, this input forms the pooling of the activities of middle temporal neurons firing according to their preferred directions. This input current is only present during the presentation of the stimulus and is shut down once the decision is made. 
            
             In the present model, in line with a large literature modeling decision making, the input, Equation~(\ref{Istim}),  is thus reduced to a signal parametrized by a scalar quantifying the coherence or degree of ambiguity of the stimulus. More global approaches consider the explicit coupling between the encoding  and the decision neural populations, with a population of stimulus-specific cells for the coding layer -- see e.g.~\cite{Beck2008,BonnasseGahot2012,Engel2015}. We believe that the main results presented here would not be affected by extending the model to take into account the coding stage, but we leave such study for further work. 
			
			In addition to the stimulus-selective part, each unit receives individually an extra noisy input, fluctuating around the mean effective external input $I_0$:
			\begin{equation}
			\tau_{noise} \dfrac{\mathrm{d} I_{noise,i}}{\mathrm{d} t} = - \left( I_{noise,i} (t) - I_0 \right) + \eta_i (t) \sqrt{\tau_{noise}} \sigma_{noise}
			\label{noise}
			\end{equation}
			with $\tau_{noise}$ a synaptic time constant which filter the (uncorrelated) white-noises, $\eta_i(t), i=L, R$. For the simulations, unless otherwise stated parameters values will be those of Table~\ref{Table1}.
			
			\begin{table}[!ht]
				
				\centering
				
				\renewcommand{\arraystretch}{1.3}
				
				\setlength{\tabcolsep}{20pt}
				
				\begin{tabular}{ll|ll}
					
					\hline
					
					{\bf Parameter} & {\bf Value} &{\bf Parameter} & {\bf Value} \\ \thickhline
					
					a & 270 Hz/nA &   $\sigma_{noise}$ & 0.02 nA \\
					
					b & 108 Hz & $\tau_{noise}$ & 2 mS\\
					
					d & 0.154 s &  $I_0$ & 0.3255 nA \\
					
					$\gamma$ & 0.641 &	$\mu_0 $ & 30 Hz \\
					
					$\tau_S$ & 100 ms &  $J_{A,ext}$ & $5.2 \times 10^{-4}$ nA/Hz   \\
					
					$J_{N,LL}=J_{N,RR}$ & 0.2609 nA & $J_{N,LR}=J_{N,RL}$ & 0.0497 nA \\
					
					$\theta$  & 20 Hz    \\
					
					\hdashline
					
					$I_{CD,max}$ & 0.035 nA &$\tau_{CD}$  & 200 ms
					
				\end{tabular}
				\caption{Numerical values of the model parameters: above the dashed line, as taken from~\cite{Wang:2006}; below the dashed line, values of the additional parameters specific to the present model (see text).}
				\label{Table1}
			\end{table}

			Initially the system is in 
            a symmetric (or neutral) attractor state,
            with low firing rates and synaptic activities (see Figure~\ref{fig1} panel B). On the presentation of the stimulus, the system evolves towards one of the 
            two 
            attractor states, corresponding to the decision state. In these attractors, the 'winning' unit fires at a higher rate than the other. We are interested in reaction time experiments. In our simulations, we consider that the system has made a decision when for the first time the firing rate of one of the two units crosses a threshold $\theta$, fixed here at $20$~Hz. We have chosen this parameter value, slightly different from the one in \cite{Wang:2006}, from the calibration of the extended model discussed below on sequential decision trials with short response-stimulus intervals (RSI). 
			We have checked that this does not affect the psychometric function of the network, 
            the accuracy is unchanged and the reaction time is shifted by a constant.

			\subsection*{Extended reduced model: inhibitory corollary discharge}
			\label{wongwangmodified}
			
			Studies~\citep{Roitman2002,Ganguli2008} show that, during decision tasks, neurons activity experiences a rapid decay following the responses - see e.g. Figures 7 and 9 in \cite{Roitman2002}. Simulations of the above model
            show that even when the stimulus is withdrawn at the time of decision, the decrease in activity is not sufficiently strong to account for these empirical findings. Decreasing the recurrent excitatory weights does allow for a stronger decrease in activity, as shown by \cite{Bonaiuto2016}. However, both the increase  and the decay of activities are too slow, and the network cannot perform sequential decisions with RSIs below 1sec. Hence the decrease in activity requires an inhibitory input at the time of the decision.

				Such inhibitory mechanism has been proposed to originate from the superior colliculus (SC), controlling  saccadic eye movements, and the basal ganglia-thalamic circuit (BG), which plays a fundamental role in many cognitive functions including perceptual decision-making. 
				Indeed, the burst neurons of the SC receive inputs from the parietal cortex and project to midbrain neurons responsible for the generation of saccadic eye movements~\citep{HallWilliamCandMoschovakis2003,Scudder2002}.
				Thus the threshold crossing of the cortical neural activity is believed to be detected by the SC~\citep{Saito2003}. In turn, the SC projects feedback connections on cortical neurons~\citep{Crapse2009}. 
				At the time of a saccade, SC neurons emit a corollary discharge  (CD) through these feedback connections~\citep{Sommer2008}. 
				The impact of this CD as an inhibition has been discussed in various contexts
				~\citep{Crapse2008,Sommer2008,Yang2008}. 
				The generation of a corollary discharge resulting in an inhibitory input has been proposed and discussed in several modelling works, in the case of the modulation of the decision threshold in reaction time tasks~\citep{Lo2006}, in the context of learning~\citep{Engel2015}, and in 
				a ring model of visual working memory~\citep{Bliss2017a}.
			
			 We note here that, for simplicity and in accordance with the existing literature~\citep{Lo2006,Engel2015,Bliss2017a}, we will be referring to the inhibitory current resulting from the corollary discharge as the {\it corollary discharge}.

			In the context of attractor networks for decision tasks, \cite{Lo2006} introduce an extension of the biophysical model of~\cite{wang2002probabilistic} consisting in modeling the coupling between the network, the  basal ganglia and the superior colliculus. The net effect is an inhibition onto the populations in charge of making the decision. 
			While \cite{Lo2006} address the issue of the control of the decision threshold, they do not discuss the relaxation dynamics induced by the corollary discharge, nor the effects on sequential decision tasks outside a learning context~\citep{LoReward}. 
			
\begin{figure}[!ht]
				\centering
				 \includegraphics[width=1.0\textwidth]{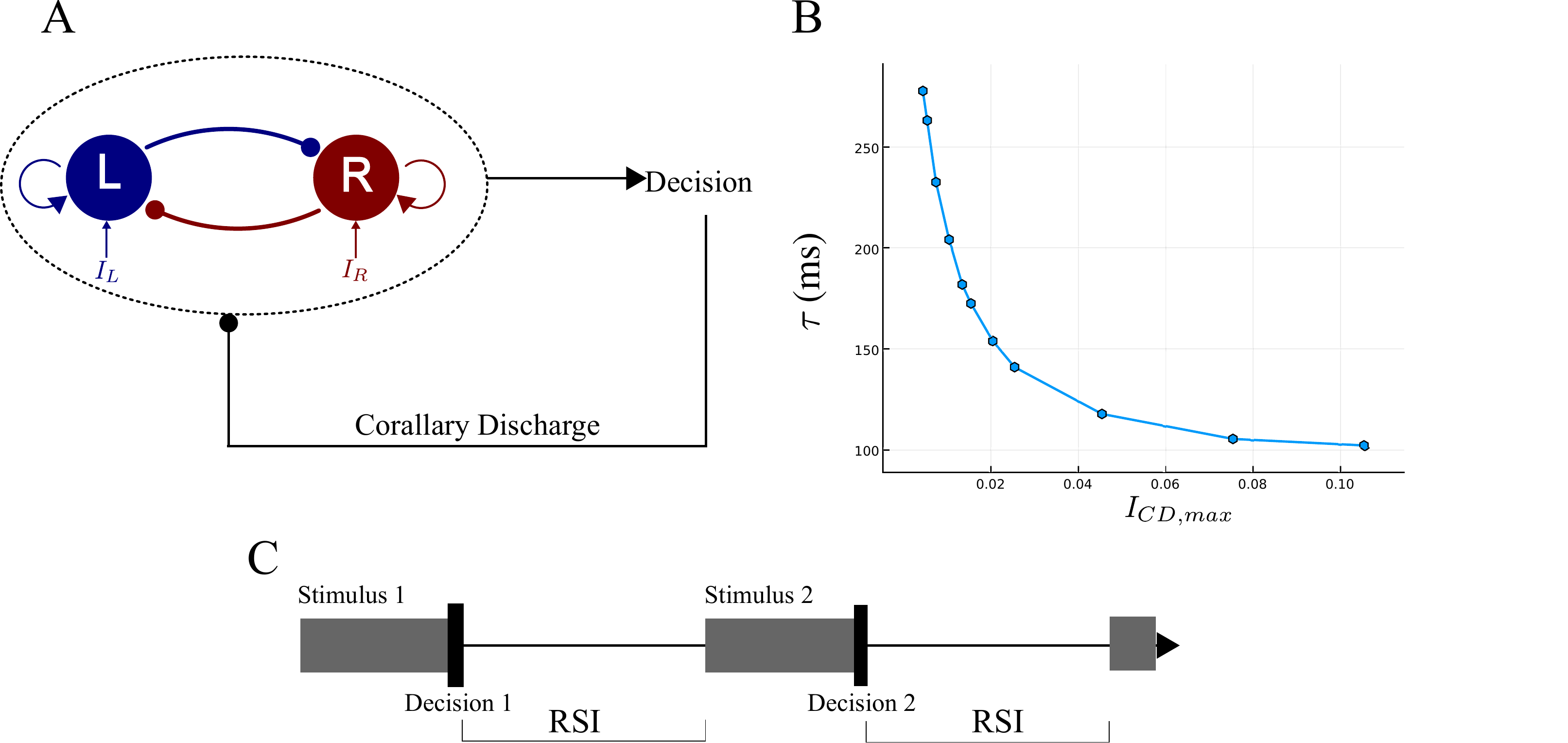} 
				\caption{{\bf Extended version of the reduced model with the corollary discharge.} 			(A)
                          The extension consists in adding                           the corollary discharge originating from the basal ganglia, an inhibitory input onto both units occurring just after a decision is made.
                           (B) Relaxation time constant of the system during the RSI (that is the relaxation dynamics towards the neutral attractor), with respect to the corollary discharge amplitude. The values are obtained by computing the 
 largest eigenvalue $\lambda$ of the dynamical system, Equation~(\ref{dSidt}--\ref{noise}),  when presenting a constant corollary discharge. The time constant is given by the inverse of the eigenvalue, $\tau = -1/\lambda$. 
                            (C) The time-sketch of the simulations can be decomposed into a succession
of identical blocks. Each block, corresponding to one trial, consists of: the presentation of a stimulus with a randomly chosen coherence (gray box), a decision immediately followed by the removal of the stimulus, a waiting 
time of constant duration corresponding to the response-stimulus interval (RSI).} 
				\label{fig2}
			\end{figure}

			In order to analyze these effects with the reduced attractor network model, we assume that, after crossing the threshold, the network receives an inhibitory current, mimicking the joint effect of basal-ganglia and superior colliculus on the two neural populations (Figure~\ref{fig2}.A). 	
            
			In the case of~\cite{Engel2015}, the function of the corollary discharge is to reset the neural activity in order to allow the network to learn during the next trial.
			For this, the form of the CD input is chosen as a constant inhibitory current for a duration of $300$ms.
			However, such strong input leads to an abrupt reset to the neural state with no memory of the previous trial. 
			We thus rather consider here a smooth version of this discharge, considering that 
			 the resulting inhibitory input has
				a standard exponential form~\citep{Finkel1983}. 
				The inhibitory input, $I_{CD}(t)$, is then given by: 
				\begin{equation}
				I_{CD}(t) = \left\{ \begin{aligned}
				& 0 & \text{ during stimulus presentation} \\  
				& - I_{CD,max} \exp \left( -(t-t_D)/\tau_{CD} \right) & \text{ after the decision time, $t_D$}   
				\end{aligned}\right.  
				\label{ICD}
				\end{equation}            
				The relaxation time constant $\tau_{CD}$ is chosen in the biological range of synaptic relaxation times and in accordance with the relaxation-times range of the network dynamics, $\tau_{CD} = 200$~ms  (see Figure~\ref{fig2}.B, and discussion below). 
                
                Therefore the input currents are modified as follows:
				\begin{eqnarray}
				I_{L,tot}(t) &= J_{LL} S_L(t) - J_{L,R} S_R(t) + I_{stim,L}(t) + I_{noise,L}(t) + I_{CD}(t) \label{ItotextL} \\
				I_{R,tot}(t) &= J_{RR} S_R(t) - J_{R,L} S_L(t) + I_{stim,R}(t) + I_{noise,R}(t) + I_{CD}(t).   
				\label{ItotextR}
				\end{eqnarray}

                We can now study the dynamics of this system in a sequence of decision trials (protocol illustrated in Figure \ref{fig2}.C). 
               We address here two issues: first, is there a parameter regime for which the network can engage in a series of trials - that is, for which the state of the dynamical system, at the end of the relaxation period (end of the RSI), is close to the neutral state (instead of being trapped in the attractor reached at the first trial);  second, is there a domain within this parameter regime for which one expects to see sequential effects (instead of a complete loss of the memory of the previous decision state).

				\begin{figure}
					\centering
					 \includegraphics[width=0.8\textwidth]{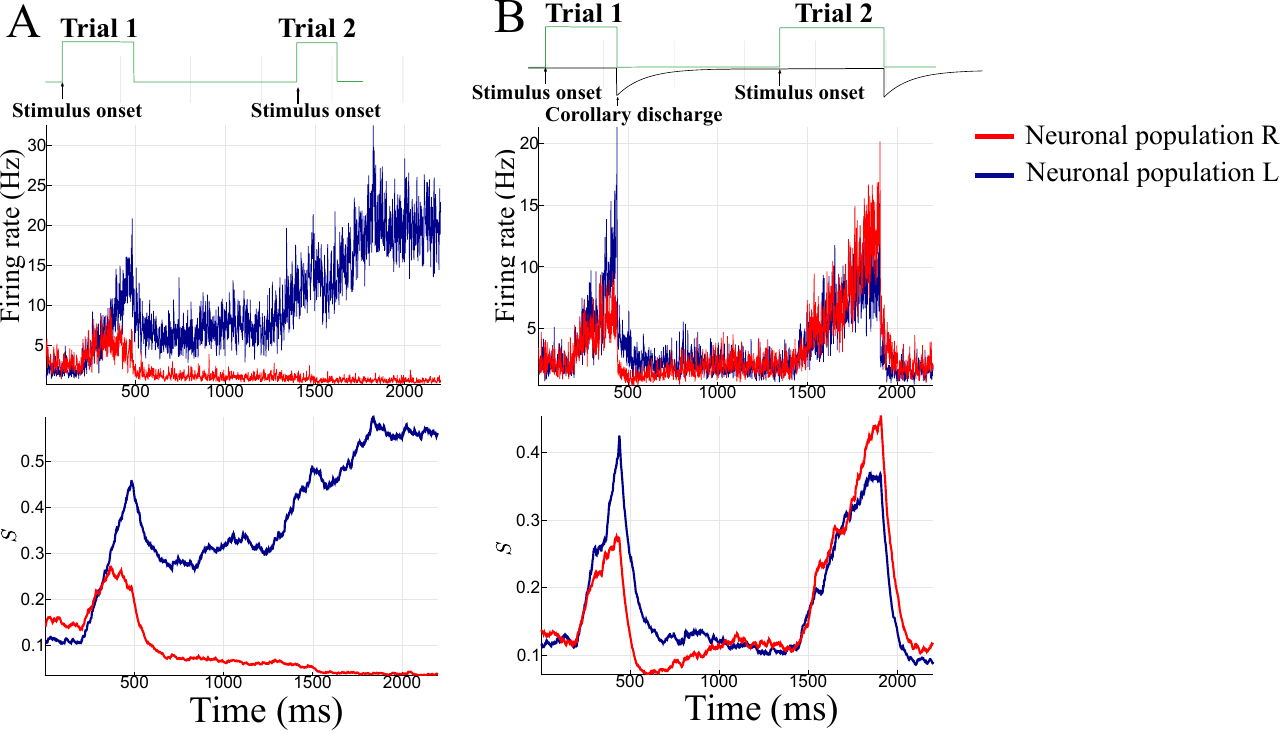} 	
					\caption{{\bf Time course of activities during two consecutive trials.} 
						Left panel, A: Without corollary discharge. 
                        A, Top (green plot): 	 Time course of the stimulations. The first stimulus belongs to category L, the second to category R.
								 A, Middle: 	 firing rates of the $L$ (blue) and $R$ (red) neural pools.  A, Bottom: corresponding synaptic activities. The neural activity becomes stuck in the attractor corresponding to the first decision. Right panel, B: With 
                            corollary discharge, with $I_{CD,max}=0.035$~nA. 	 B, Top:   Time course of the stimulations (green plot, same protocol as for (A)), and time course of the inhibitory current (black curve, represented inverted for clarity of the presentation). B, Middle and Bottom: neural and synaptic activities, respectively (L pool: Blue, R pool: Red). In that case, one observes the decay of activity after a decision has been made, and the winning population is different for the two trials.					
					}
					\label{fig3}

				\end{figure}

				In Figure~\ref{fig3} we illustrate the network dynamics between two consecutive stimuli during a sequence of trials, comparing the cases with and without the corollary discharge.  
				In the absence of the CD input, the network is not able to make a new decision different from the previous one (Figure~\ref{fig3}.A). Even when the opposite stimulus is presented, the system cannot leave the attractor previously reached, unless in the presence of an unrealistic strong input bias.     
				If however the strength $I_{CD,max}$ is strong enough, the corollary discharge makes the system to escape from the previous attractor and to relax towards near the neutral resting state with low firing rates. If too strong, or in case of a too long RSI, at the onset of the next stimulus the neutral state has been reached and memory of past trials is lost. For an intermediate range of parameters, at the onset of the next stimulus the system has escaped from the attractor but is still on a trajectory dependent on the previous trial (Figure~\ref{fig3}.B).

                We have computed the time constant $\tau$ of the network during relaxation (during the RSI), with respect to the CD amplitude, $I_{CD,max}$, see Figure~\ref{fig2}.B. This computation is done for a  corollary discharge with a constant amplitude, $I_{CD}(t)=I_{CD,max}$. 
             One sees that, for $I_{CD,max}$ of order $0.03\sim 0.04$nA, the network time constant $\tau$ is four to  five times smaller than the duration of the RSI. We choose the relaxation constant $\tau_{CD}$ of the corollary discharge of the same order of magnitude (as in the above simulation where $\tau_{CD}=200$ms). With such value, at the onset of the next stimulus, the network state will still be far enough from the symmetric attractor, so that we can expect to observe sequential effects, as confirmed by the analysis in the Results Section.

				With the inhibitory corollary discharge, after the threshold is crossed by one of the two neural populations, there is  a big drop in the neuronal activity (Figure~\ref{fig3}.B), corresponding to the exit from the previous attractor state. 
					This type of time-course is in agreement with the experimental findings of~\cite{Roitman2002,Ganguli2008}, who measure the activity of LIP neurons during a decision task. 
					They show that neurons that accumulate evidence during decision
					tasks experience rapid decay, or inhibitory suppression, of activity following responses, similar to Figure~\ref{fig3}.B 
					(but see~\cite{Lo2006} for a related modeling study with spiking neurons, or \cite{Gao2009} for rapid decay of neural activity with an other type of attractor network).
                 
\begin{figure}
					\centering
					\includegraphics[width=1\textwidth]{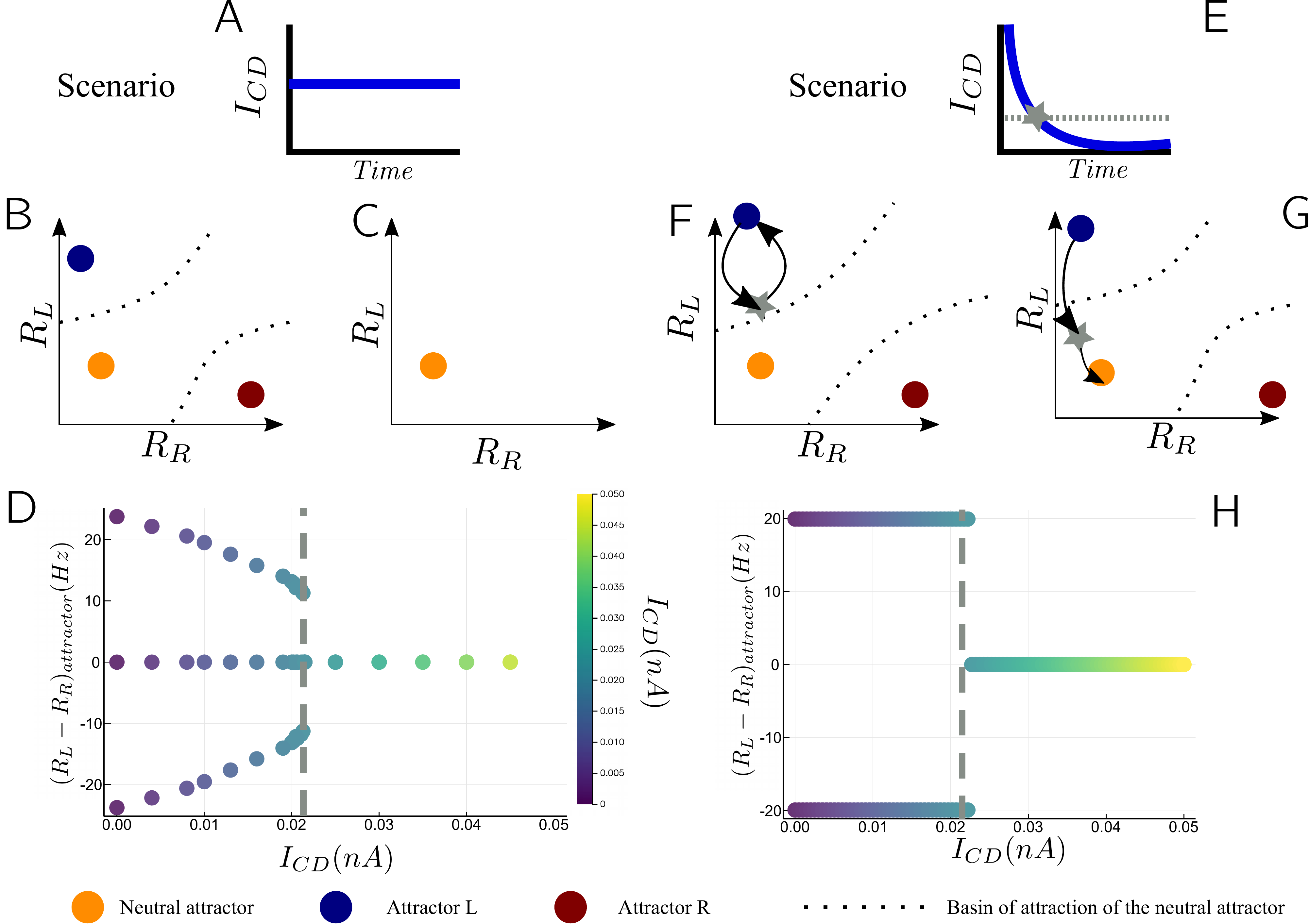}  
					\caption{     {\bf Bifurcation diagram of sequential decision making, for two scenario of $I_{CD}$.}
							  (A) Scenario with a constant value of the inhibitory current for the left part of the figure, panels B, C and D.
                                                        (B) Phase plane representation of the attractors at low $I_{CD}$ (below the critical value). 
                            							(C) Phase plane representation of the attractor landscape at high $I_{CD}$ (above the critical value). Only the neutral attractor exists, corresponding to the right side of panel D.
                                                        	(D) Attractors state (as the difference in firing rates, $R_L - R_R$) with respect to $I_{CD}$. The gray line, at $I_{CD} = 0.0215$ nA, represents the bifurcation point. On the left side three attractors exists, on the right side only the neutral one exists. The case without inhibitory current corresponds to $I_{CD}=0$ nA.
							(E) Scenario with an inhibitory current decreasing exponentially with time, for the right part of the figure, panels F, G and H.							The dashed line corresponds to $I_{CD}=0.0215$ nA, value at which the bifurcation at constant $I_{CD}$ occurs (see panel D). The time at which the current amplitude crosses this value is denoted by the gray star in panels E and F.
														(F) Schematic phase-plan dynamics corresponding to the left side of (H). The blue attractor corresponds to the starting point and the black arrow represents the dynamics. At the time $I_{CD}$ becomes lower than $0.0215$ nA (gray star), the system is still within the basin of attraction of the initial attractor. Hence, it goes back to the initial attractor.
							(G) Schematic phase-plan dynamics corresponding to the right side of panel H. At the time $I_{CD}$ becomes lower than $0.0215$ nA, the system lies within the basin of attraction of the neutral attractor. Hence, the dynamics continues towards the neutral attractor. Those conditions are the ones needed for sequential decision-making.
													(H) Attractors that can be reached when starting from a decision state, for the relaxation dynamics under the scenario represented on panel E.
							On the left side of the dashed gray line, the value of $I_{CD,max}$ is too weak and the network remains locked to the attractor corresponding to the previous  decision state. On the right side the network dynamics lies within the basin of attraction of the neutral attractor, allowing the network to engage in a new decision task. 		 }
					\label{fig4}					
				\end{figure}

                  We now derive the conditions 
                 on $I_{CD}$ 
                 under which the network is able to make a sequence of trials. To this end, we analyze the dynamics after a decision has been made, during the RSI (hence during the period with no external excitatory inputs). 
              The results are illustrated in Figure~\ref{fig4} on which we represent a sketch of the phase plane dynamics and a bifurcation diagram.

                 Consider first what would happen under a scenario of a constant, time independent, inhibitory input during all the RSI (Figure~\ref{fig4}.A-B-C-D) (formally, this correspond to setting $\tau_{CD}=+\infty$ in Equation~\ref{ICD}). 
                At small values of the inhibitory current, the attractor landscape is qualitatively the same as in the absence of inhibitory current:   
                in the absence of noise there is three fixed points, one associated with each category and the neutral one (Figure~\ref{fig3}.B). At some critical value, of about $0.0215$ nA, there is a bifurcation (Figure~\ref{fig4}.D): for larger values of the inhibitory current, only one fixed point remains, the neutral one (Figure~\ref{fig4}.D). As a result, applying a constant CD would either have no effect on the attractor landscape - current amplitude below the critical value - so that the dynamics remains within the basin of attraction of the attractor reaches at the previous trial; or would reset the activity at the neutral state (current amplitude above the critical value), loosing all memory of the previous decision. 
					
					Now in the case of a CD with a value decreasing with time (Figure~\ref{fig4}.E-F-G-H, scenario of an exponential decay), the network behavior will depend on where the dynamics lies at the time of the onset of the next stimulus. 
The dynamics, starting from a decision state (e.g. near the  blue attractor in Figure~\ref{fig4}.F-G),  is more easily understood by considering the limit of slow relaxation (large time constant $\tau_{CD}$). Between times $t$ and $t+\Delta t$, with $\Delta t$ small compared to $\tau$, the dynamics is similar to what it would be with a constant CD with amplitude $I_{CD}(t)$. Hence if $I_{CD}(t)$ is larger than the critical value discussed above, the dynamics 'sees' a unique attractor, the neutral state, and is driven towards it. When $I_{CD}(t)$ becomes smaller than the critical value, the system 'sees' again three attractors, and finds itself within the basin of attraction of either the initial fixed point (corresponding to the previous decision, Figure~\ref{fig4}.F), or of the neutral fixed point (Figure~\ref{fig4}.G). 
					In the latter case, the network is able to engage in a new decision task.

					To conclude, to have the network performing sequential decision tasks, one needs $I_{CD,max}$ to be larger than the critical value (about $I_{CD}=0.0215$ nA, Figure~\ref{fig4}.H), and, for a given value of $I_{CD,max}$, to have a time constant $\tau_{CD}$ large enough compared to the RSI for the system to relax close enough to the neutral attractor at the onset of the next stimulus.  However, sequential effects may exist only if the current decreases sufficiently rapidly, so that the trajectory is still significantly dependent on the state at the previous decision. This justifies the choice of exponential decrease of the inhibitory current, Equation~\ref{ICD}, and the numerical value of $\tau_{CD}=200$ms.   We note that recording from relay neurons, \cite{Sommer2002}   show that the signal corresponding to the corollary discharge last several hundred of milliseconds. This time scale falls precisely in the range of values 
            of the relaxation time constant of the model  (see Figure~\ref{fig2}.B),
            and corresponds to values for which, as we will see, the model shows sequential effects.

				\subsubsection*{Numerical simulations design and Statistical tests}
				\label{app:simulations}
				
				\paragraph{Numerical simulations}$\;$\\ 	
                               The simulation of sequential decision-making is as follows: a stimulus with a randomly chosen coherence is presented until the network reaches a decision 
                (decision threshold crossed). The decision is immediately followed by the removal of the stimulus, and 
                a relaxation period during the response-stimulus interval (RSI). Then a new stimulus is presented, initiating the next trial (Fig.~\ref{fig2}.C). 
				The set of dynamical equations (\ref{dSidt},\ref{noise}) -- with the definitions (\ref{fI},\ref{Istim},\ref{ICD},\ref{ItotextL},\ref{ItotextR}) -- 
				is numerically integrated using Euler-Maruyama method with a time step of $0.5$~ms. At the beginning of a simulation, the system is set in a symmetric state $S_L=S_R=s_0$, with low firing rates and synaptic activities, $s_0=0.1$. We compute the instantaneous population firing rates, or the synaptic dynamical variables $S_L$ and $S_R$, by averaging on a time window of $2$~ms, slided with a time step of $1$~ms. The accuracy of the network's performance is defined as the percentage of trials for which the units crossing the threshold corresponds to the stronger input.
				For data analysis we mainly work with the variables $S_L$ and $S_R$,  which are analog to the firing rates $R_L$ and $R_R$ (since they are monotonic function of $S_L$ and $S_R$, \cite{Wang:2006}),
                but are  less noisy (see Fig~\ref{fig3}). 
				We consider that the system has made a decision when for the first time the firing rate of one unit crosses a threshold $\theta$, fixed at $20$~Hz.
				The reaction time during one trial is defined as the time needed for the network to reach the threshold from the onset of the input stimulus. We neglect the possible additional time due to motor reaction and signal transduction.
                In addition to the reaction times, we compute the {\it discrimination threshold}, which is linked to the accuracy. The definition is based on the use of a Weibull function commonly used to fit the psychometric curves~\citep{Quick1974}. That is, one writes the performance (mean success rate) as $\text{Perf}(c) = 1 - 0.5 \exp\left(- \left(c/\alpha\right)^{\beta} \right)$, where $\alpha$ and $\beta$ are parameters. Then, for $c=\alpha$, $\text{Perf}(c)= 1-0.5 \exp\left(-1\right) \sim 0.82$. Hence one defines the discrimination threshold as the coherence level at which the subject responds correctly $82 \%$ of the time.

				We list in Table~\ref{Table1} the model parameters which correspond to the one of the simulations. For Figures~\ref{fig5} and \ref{fig7} we have used continuous sequences of $1000$ trials averaged over $24$ independent simulations   - allowing to more specifically compare with experiments of~\cite{Bonaiuto2016} done with $24$ subjects. Figures 9 to 16 present results obtained for sequences of $1000$ trials averaged over $50$ independent simulations to allow for a better statistical analysis. The number of sequences, $1000$, is a typical order of magnitude in experiments (see e.g.~\cite{Bonaiuto2016} and~\cite{Danielmeier2011}).

				\paragraph{Statistical tests}$\;$\\ Following~\cite{Benjamin}, we consider a p-value of $0.005$ as a criterion for rejecting the null hypothesis in a statistical test. To assess if the distributions of two continuous variables are different, we make use of the Kolmogorov-Smirnov test~\citep{Hollander2014}, and in the case of discrete variable distributions we use the Anderson-Darling test~\citep{Shorack2009}. For very large samples, we use the energy distance~\citep{Rizzo2016}, which is a metric distance between the distributions of random vectors. We use the associated E-statistic~\citep{Szekely2013} for testing the null hypothesis that two random variables $X$ and $Y$ have the same cumulative  distribution functions. 
				For testing whether the means of two samples are different we make use of the Unequal Variance test (Welch's test)~\citep{Hollander2014}.
				
				\paragraph{Softwares and Code accessibility}$\;$\\ 
				For the simulations we made use of the Julia language~\citep{Bezanson2014Julia:Computing}.  
				The code of the simulations can be obtained from the corresponding author upon request.
				We made use of the XPP software~\citep{Ermentrout2003} for the phase-space analysis and the computation of the relaxation time constant of the dynamical system. Figures $9,10,11,12$ and $19$ were realized using Python and the other are in the same language as the simulations. The E-statistics tests were performed using the R-Package: {\it energy package}~\citep{Rizzo2014}.
				
				\section*{Results}
				
				\subsection*{Sequential dynamics and choice repetition biases}
				\label{sec:repetition}

                The dynamical properties described above give that, for the appropriate parameter regime, the RSI relaxation leads to a state which is between the previous decision state and the neutral attractor. If it is still within the basin of attraction of the previous decision state at the onset of the next stimulus,  one expects sequential biases. This mechanism is similar to the one discussed by \cite{Bonaiuto2016}. However, the relaxation mechanisms are different, as discussed in the Introduction.  This results in different quantitative properties, notably and quite importantly  in the time scale of the relaxation, which is here more in agreement with experimental findings~\citep{Cho2002}. 
                
                We will specifically show that nonlinear dynamical effects are at the core of post-error adjustments. As a preliminary step, it is necessary to investigate the occurrence of sequential effects in our model. We do so by describing more precisely the inter-trial dynamics: we need to specify where the network state lies at the onset of a new stimulus, with respect to the boundaries between the basins of attraction. We take advantage of this analysis to explore response repetition bias as studied in \cite{Bonaiuto2016}, and to confront the model behavior with other empirical findings~\citep{Laming1979,Cho2002}. In all the following, we  study the model properties in function of the  two parameters, the amplitude of the corollary discharge, $I_{CD,max}$, and the duration of the RSI.

				\subsubsection*{Network behavior: Reaction times biases}
				 After running simulations of the network dynamics with the protocol of Figure~\ref{fig2}.C, we analyze the effects of response repetition by separating the trials into two groups, the {\it Repeated} and {\it Alternated} cases. The repeated (respectively alternated) trials are those for which the decision is identical to (resp. different from) the decision at the previous trial.     
								Note that we do not consider whether the {\it stimulus} category is repeated or alternated: the analysis is based on whether the {\it decision}                                 is identical or different between two consecutive trials~\citep{Fleming2010,Padoa-Schioppa2013}.  Such analysis is appropriate, since the effects under consideration depend on the levels of activity specific to the previous decision.  				
				We run a simulation of $1 000$ consecutive trials, each of them with a coherence value randomly chosen between $20$ values in the range  $[-0.512, 0.512]$.  We do so for two values of the corollary discharge amplitude, $I_{CD,max}=0.035$~nA and $I_{CD,max}=0.08$~nA, with a RSI of $1$~s, the other parameters being given on Table~\ref{Table1}.
				
\begin{figure}
					\centering
					 \includegraphics[width=0.9\textwidth]{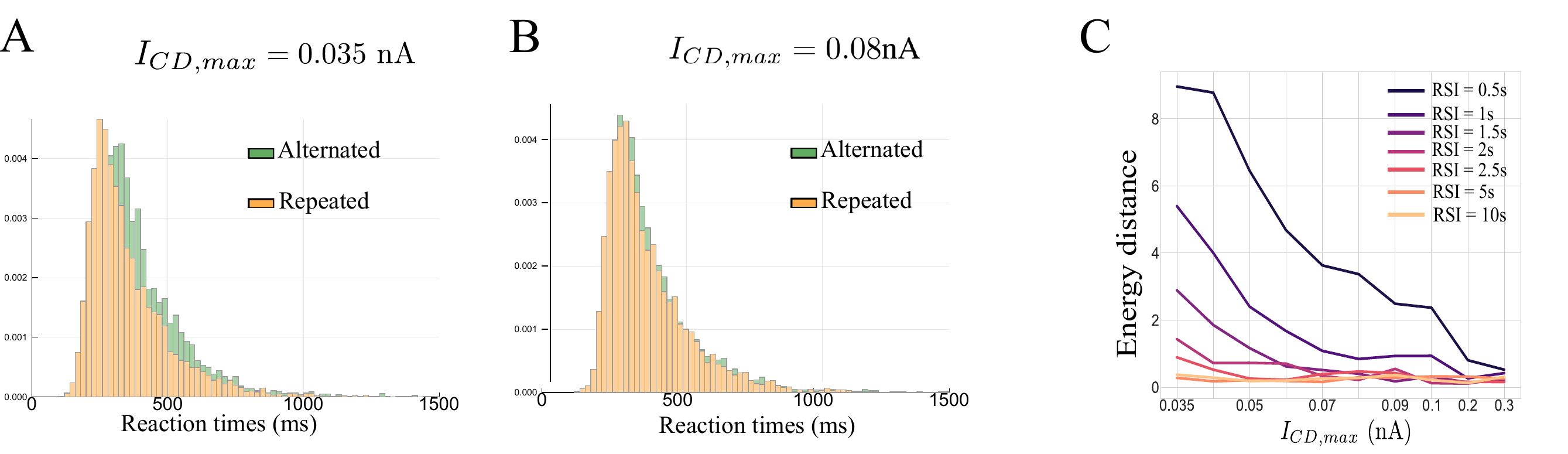} 
					\caption{{\bf Histogram of the reaction times.}
						Simulations run at,  (A): $I_{CD,max}=0.035$~nA, and (B): $I_{CD,max}=0.08$~nA, with a RSI of $1$~second. The green histogram  corresponds to the Alternated case, that is when the decisions made at the $n$th and $n$th~$+1$ trials are different. The orange histogram corresponds to the Repeated case, that is when the decisions made at the $n$th and $n$th~$+1$ trials are identical.  (C) Energy distance between the repeated and alternated histograms. The $x$-axis represents the strength of the corollary discharge, and the color codes the duration of the RSI in seconds.     
					}
					\label{fig5}
				\end{figure}

				We find that the distribution of coherence values are identical for the two groups, for both values of $I_{CD,max}$ (Anderson-Darling test, $p=0.75$ and $p=0.84$ respectively). 
				We study the reaction times separately for the two groups, and present the results in Figure~\ref{fig5}.
				In Figure~\ref{fig5}.C we represent the so called energy distance~\citep{Szekely2013,Rizzo2016} between the repeated and alternated reaction times distribution. As we can observe, the distance decreases, hence the sequential effect diminishes, as the corollary discharge amplitude $I_{CD,max}$ increases. 
				For the specific case of Figures~\ref{fig5}.A and B, the corresponding E-statistic for testing equal distributions leads to the conclusion that in the case $I_{CD,max} = 0.035$~nA, the two reaction-time distributions are different ($p=0.0019$). This implies that the behavior of the network is influenced by the previous trial. We observe a faster mean reaction time (around $55$~ms) when the choice is repeated (Figure \ref{fig5}A), with identical shape of the reaction times distributions. The difference in means is of the same order as found by \cite{Cho2002} in experiments on 2AFC tasks. 
               On the contrary, for $I_{CD,max} = 0.08$~nA  (Figure~\ref{fig5}.B), the two histograms cannot be distinguished (E-statistic test, $p=0.25$).
				
				We have checked that increasing the RSI has a similar effect to increasing the corollary discharge amplitude. We observe sequential effects for RSI values in the range $0.5$ to $5$~seconds, in accordance with two-choices decision-making experiments, where such effects are observed for RSI less than $5$ seconds~\citep{Rabbitt1977,Laming1979,Soetens1985}.

				\subsubsection*{Neural correlates: Dynamics analysis}
				With the relaxation of the activities induced by the corollary discharge, the state of the network at the onset of the next stimulus lies in-between the attractor state corresponding to the previous decision, and the neutral attractor state. 
				When averaging separately over repeated and alternated trials, we find, as detailed below, that this relaxation dynamics has different behaviors depending on whether the next decision is identical or different from the previous one. Note that this is a statistical effect which can only be seen by averaging over a very large number of trials. 
				
				\begin{figure}
					\centering
					\includegraphics[width=0.9\textwidth]{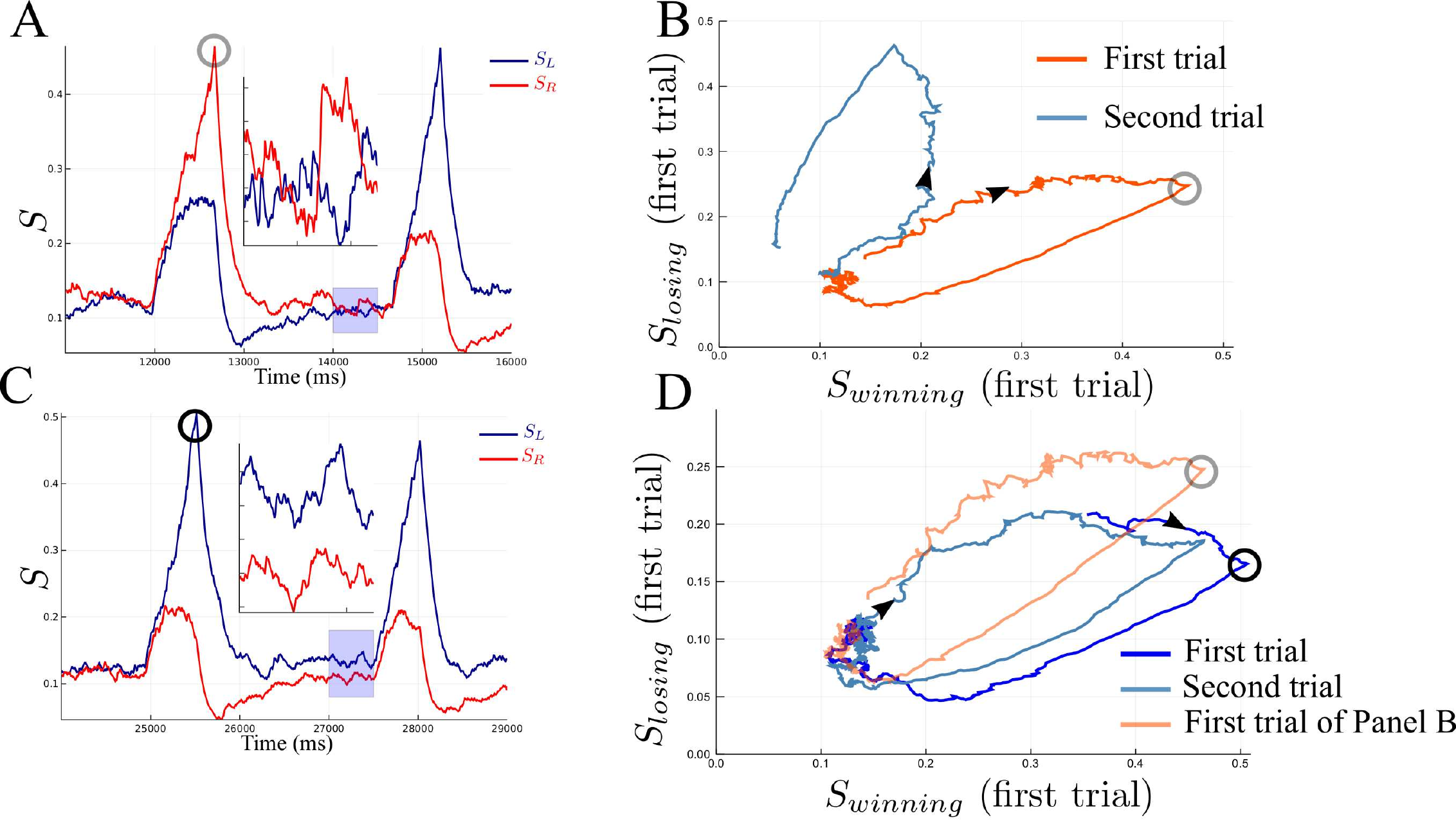} 
					\caption{{\bf Network activity during two consecutive trials.}  
						Panels (A) and (B) represent the alternated case where the decision made is $R$ then $L$, and panels (C) and (D) represent the case where decision $L$ is made and repeated. Panels (A) and (C) plot the time course activities of the network. The light blue zone is zoomed in order to better see the dynamics just before the onset of the second stimulus. The red and blue curves correspond to the activities of, respectively, the $R$ and $L$ network units. 
						Panels (B) and (D) represent, respectively, the (A) and (C) dynamics in the phase-plane coordinates. On panel (B) the dynamics evolves from dark red (first trial)  to light blue (2nd trial), and on panel (D) from dark blue (first trial) toward light blue (2nd trial). The gray -- respectively black -- circles identify the same specific point during the dynamics in panels (A) and (B) -- resp. (C) and (D). 
							The circles are not at the exact same value because the decision threshold is on the firing rates and not for the synaptic activities.
						In order to compare the alternated and repeated cases, (A,B) and (C, D),  the dark red curve of panel (B), is reproduced on panel (D) as light orange curve. 
					}
					\label{fig6}
				\end{figure}
				
				In Figure~\ref{fig6} we compare two examples of network activity, one with an alternated choice, and one with a repeated choice, by plotting the dynamics during two consecutive trials. We observe in Figure \ref{fig6}.A, the alternated case, that previous to the onset of the second stimulus (light blue rectangle) the activities of the two populations are at very similar levels. In contrast, for the case of a repeated choice, Figure~\ref{fig6}.C, the activities are well separated, with higher firing rates.

					In Figure~\ref{fig6}.B we give a classical phase-plane representation of the network dynamics during two consecutive trials, with the axes as the synaptic activities of the wining versus loosing neuronal populations in the first trial. One sees a trajectory starting from the neutral state, going to the vicinity of the attractor 
                    corresponding to the first decision, and then relaxing to the vicinity of the neutral state  (as illustrated in Figure~\ref{fig4}.G). 
                    Then the trajectory goes towards the attractor corresponding to the next decision, different from the first one. This aspect of the dynamics is similar to what is obtained in~\cite{Gao2009} with another type of attractor network.
				We show in Figure~\ref{fig6}.D the phase-plane dynamics in the case of a repeated choice (trajectory in blue). On this same panel, for comparison we reproduce in light red the dynamics, shown in Figure~\ref{fig6}.B, during the first trial in the alternated case. As can be seen in Figure \ref{fig6}.D, the network states at the time of decision are different depending on whether the network makes a decision identical to, or different from, the one made at the previous trial.
				
\begin{figure}
					\centering
					\includegraphics[width=0.9\textwidth]{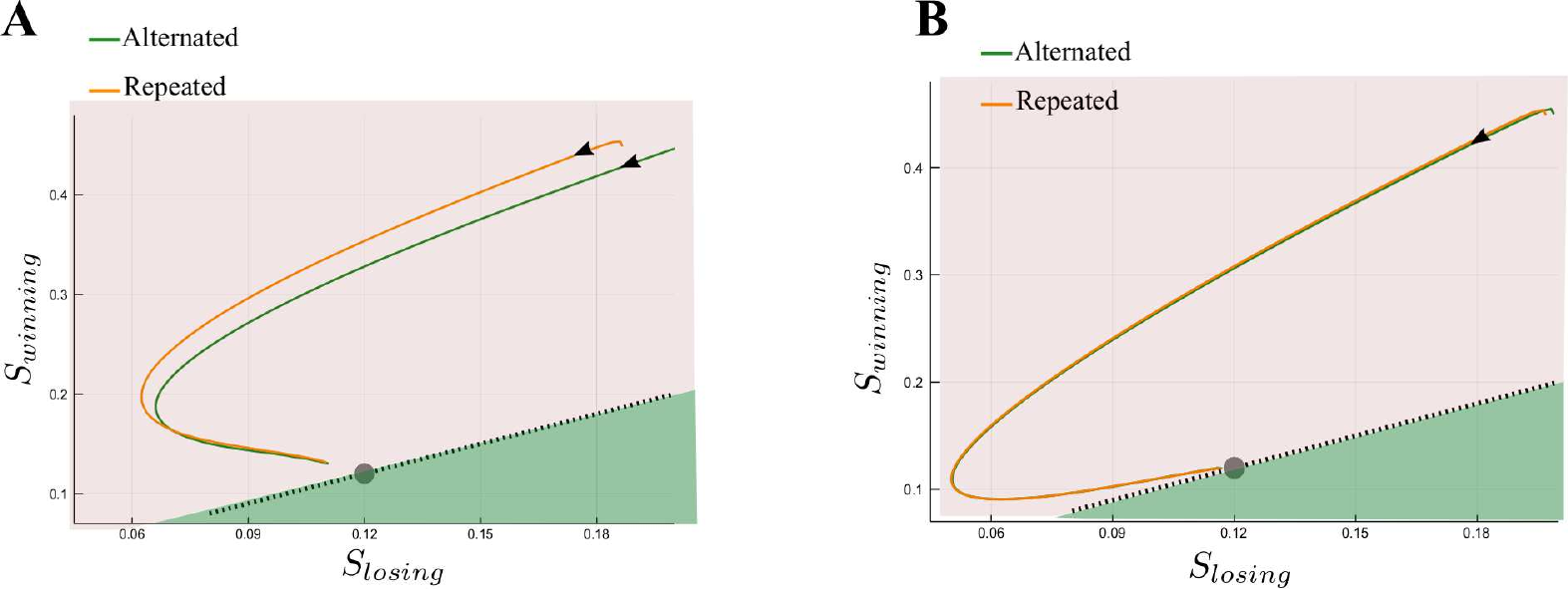}
					\caption{{\bf Phase plane dynamics. }
						Dynamics of the decaying activity between two successive trials, (A) for $I_{CD,max} = 0.035$~nA, and (B) for $I_{CD,max} = 0.08$~nA. The synaptic activity is averaged over all trials separately for each one of the two groups: alternated (green) and repeated (orange). The axis are $S_{winning}$ and $S_{losing}$ (not $S_R$ and $S_L$) corresponding to the mean synaptic activity of, respectively, the winning and the losing populations for this trial. Note the difference in scale of the two axes. The time evolution along each curve follows the black arrow. The dashed black line denotes the symmetric states ($S_L=S_R$) of the network, and the gray circle the neutral attractor. The shadow areas represent the basins of attraction (at low coherence levels) for the repeated and alternated trials, respectively pink and green. 
					}
					\label{fig7}
				\end{figure}

				In order to check the statistical significance of these observations, we represent in Figure~\ref{fig7} the mean activities during the RSI, obtained by averaging the dynamics over all trials, separately for the alternated and repeated groups. As expected, for small values of $I_{CD,max}$ ($0.035$~nA), the two dynamics are clearly different. This difference diminishes during relaxation. However at the onset of the next stimulus we can still observe some residues, statistically significant according to an Anderson-Darling test done on the $500$~ms prior to the next stimulus (between winning population, $p=0.0034$, between losing population $p=3.2 \times 10^{-8}$).

                Looking at Figure~\ref{fig7}.A, we observe that the ending points of the alternated and repeated relaxations are biased with respect to the symmetric state. At the beginning of the next stimulus the network is already in the basin of attraction of the repeated case. Hence, it will be harder to reach the alternated attractor stated (in the green region). When increasing $I_{CD,max}$ (Figure~\ref{fig7}.B), we observe that the ending state of the relaxation is closer to the attractor state. Hence, the biases in sequential effects disappear because at the beginning of the next stimuli the network starts from the symmetric (neutral) state.
				 The same analysis holds for
				longer RSI, the dynamics are almost identical (Anderson-Darling test: between winning population, $p=0.25$, between losing population $p=0.4$), and both relaxation end near the neutral attractor state. The bias depending on the next stimulus is not observed anymore, and the sequential effect on reaction time hence disappears.
                
                Note that the sequential effects only depend on whether or not the states at the end of the relaxation lie on the basin boundary. 
				However, we have just seen that the effects can also be observed at the level of the relaxation dynamics, since the trajectories for alternated and repeated cases are identical when there is no effect, and different in the case of sequential effects.
                
\begin{figure}[!ht]
						\centering
							\includegraphics[width=1.0\textwidth]{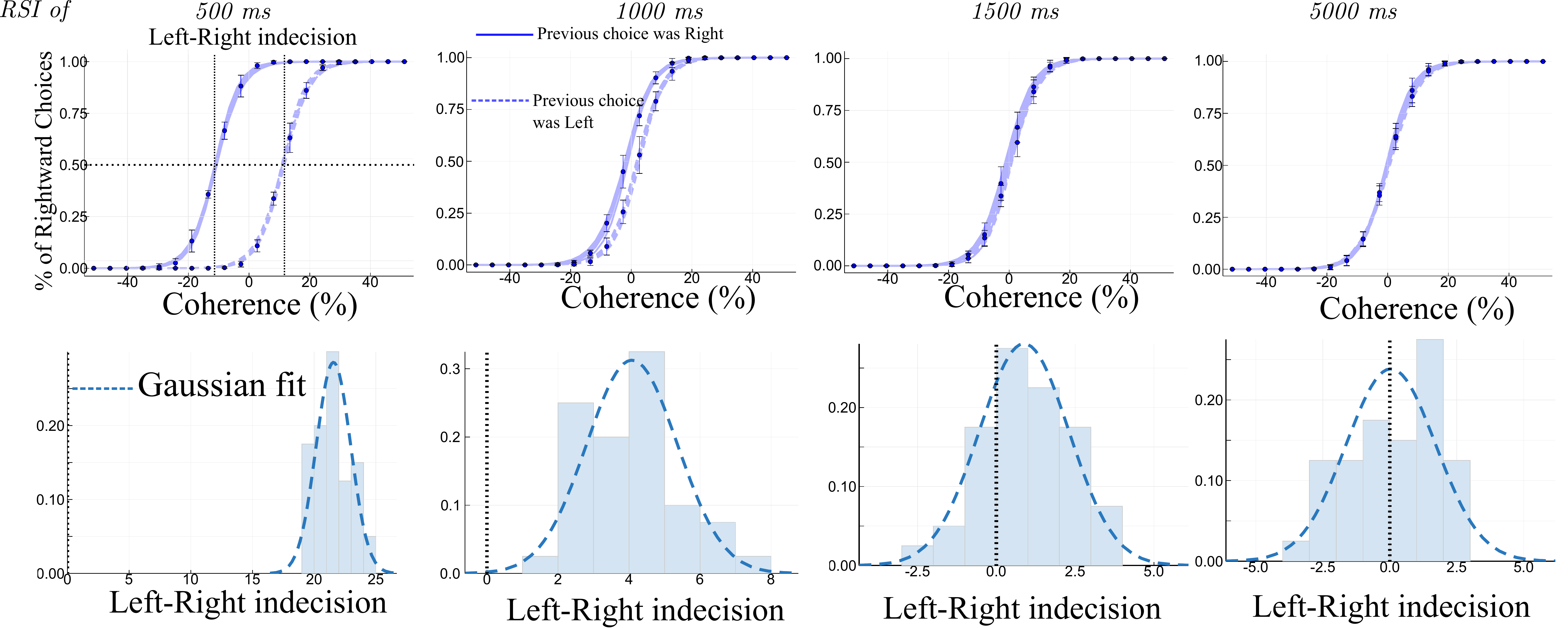} 
						\caption{	 {\bf Repetition biases for several RSI values. } 
							Upper panel: percentage of Right choices, with respect to coherence level, depending on the previous choice (Left or Right). The blue points represents the mean accuracy (on 24 simulations) and the confidence interval at $95\%$ (bootstrap method). The blue lines denote the fit (of all simulations) by a logistic regression of all  (plain: previous choice was Right, dashed: previous choice was Left). Bottom panel: histogram of the Left-Right indecision point (on 24 simulations to stay in the experimental range). It characterizes the fact that the positive shift in the indecision point is increased for small RSI. The mean of the indecision point shift decreases with longer response-stimulus intervals.}
						\label{fig8}
					\end{figure}

					The analysis of the dynamics also leads to expectations for what concerns the bias in accuracy towards the previous decision.  
					Indeed, this can be deduced from Figure~\ref{fig7}. If the choice at the previous trial
					was $R$ (respectively $L$), then, at the end of the relaxation, the network lies closer to the basin of attraction of attractor $R$ (respectively $L$). Thus when presenting the next stimulus, the decision will be biased towards the previous state, so that the probability of making the same choice will be greater than the one of making the opposite choice. 
					Otherwise stated, given the stimulus presented at the current trial, the probability to make the choice $R$ will be greater when the previous choice was also $R$, than when the previous choice was $L$. Numerical simulations confirm this analysis, as illustrated on Figure~\ref{fig8}. The RSI dependency is statistically significant 
					(Generalized Linear Model: $r=-3.9$, $p<0.0001$). For small RSI (500 ms), the decision is biased towards the previous one, and for RSI of several seconds this effect disappears. These results are in agreement with experimental findings of~\cite{Bonaiuto2016}.   The authors studied response repetition biases in human with RSIs of at least 1.5 seconds. In these  experiments, they measure the Left-Right indecision point, that is the level of coherence resulting in chance selection. Compared to the repeated case, they find that the indecision point for the alternated case is at a higher coherence level,  and this shift decreases as the RSI increases.  	    
			
			 Sequential decision effects have also been analyzed within the DDM framework~\citep{Farrell2008,Goldfarb2012}.
					Behavioral data can be fitted by different choices of starting points, and possibly of thresholds~\citep{Goldfarb2012}. 
					The modification of the starting point in a DDM framework is analog to the effect of the relaxation in our model. However, most works based on DDM make a post-hoc analysis of empirical data, with separate fits for alternated and repeated cases.

				To conclude this section, at the time of decision, the winning population has a firing rate higher than the losing population. After relaxation, at the onset of the next stimulus, the two neural pools have more similar activities, but are still sufficiently different, that is the dynamics is still significantly away from the neutral attractor. 
				At the onset of the next stimulus, the systems finds itself in the basin of attraction of the attractor associated to the same decision as the previous one.  This results in a dynamical bias in favor of  the previous decision. 
				 The probability to make the same choice as the previous one is then larger than the one of the other choice, and the reaction time, for making the same choice (repeated case), is shorter than for making the opposite choice (alternated case).
					In accordance with these results, studies on the LIP, superior colliculus and basal ganglia have found that the baseline activities before the onset of the stimuli can reflect the probabilities of making the saccade, under specific conditions~\citep{Lauwereyns2002,Ding2006,Rao2012}. Our model shows that these modulations of the baseline activities can be understood as resulting from the across-trial dynamics of the decision process.

				\subsection*{Post-error effects}
				\label{sec:post-error}
				
				\subsubsection*{Post-error adjustments on reaction times}
				The most interesting and well established effect is the one of Post-Error Slowing (PES)~\citep{Laming1979}, and see \cite{Danielmeier2011} for a review. It consists of prolonged reaction times in trials following an error, compared to reaction times following a correct trial. This effect has been observed in a variety of tasks: categorization \citep{Jentzsch2009}, flanker \citep{Debener2005}, Stroop \citep{Gehring2001} tasks.  
				\cite{Jentzsch2009} and \cite{Danielmeier2011} found that the PES effect depends on the response-stimulus interval. The amplitude of this effect, 
				defined as the difference between the mean reaction times  of post-error and post-correct trials, decreases as one increases the RSI, with values going from several dozens of milliseconds to zero. For RSI longer than $750-1500$~ms, PES is not observed anymore.  Remarkably, the PES effect is reported in cases where the subject does not receive information on the correctness of the decision~\citep{Jentzsch2009,Danielmeier_etal2011,Danielmeier2011}. 
                Moreover, this effect is automatic and involuntary \citep{Rabbitt2002}, and is independent of error detection and correction process which involve other cortical areas \citep{Rodriguez2002}. This suggests a rather low level processing origin in line with the present model.

				In this section we investigate the occurrence of post-error adjustments in our model. 
				We confront the results to empirical findings from various behavioral  experiments 
                with Two Alternative Forced Choice (and marginally also 4-AFC)  protocols
                in which, as it is also the case in our model, there is no feedback on the correctness of the decision. 
                We will notably discuss the model predictions comparing the results with those of \cite{Danielmeier2011} who  studied  the dependence of PES with respect to the RSI, as well as the relation between PES and PIA. 
 
				\begin{figure}
					\centering
					\includegraphics[width=0.9\textwidth]{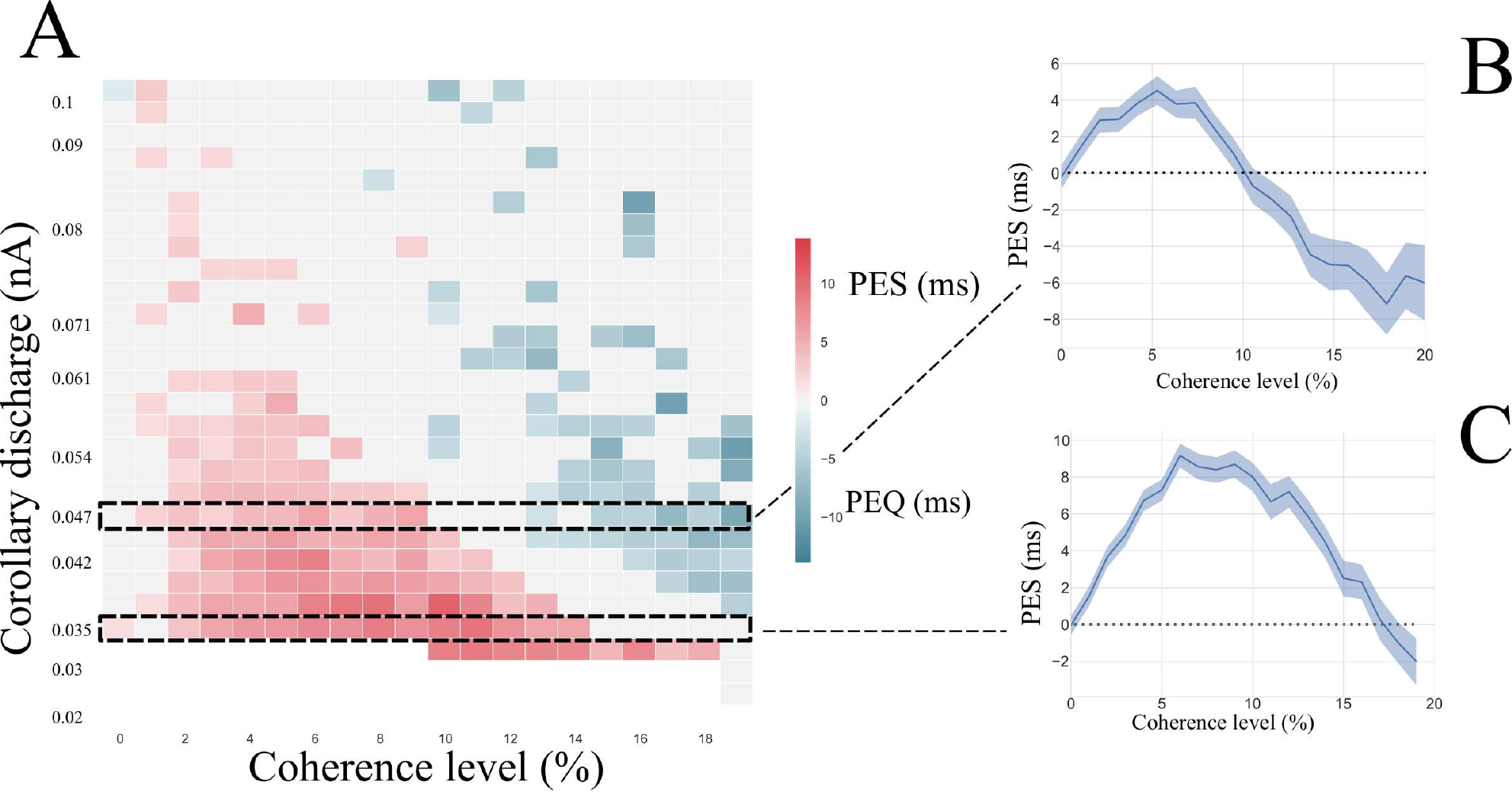}
					\caption{{\bf Post-error slowing 
							 in the simulated network at a RSI of 500 ms.					} (A) Phase diagram of the PES effect at RSI of $500$~ms. The bottom white zone corresponds to parameters where sequential decision-making is impossible as the network is unable to leave the attractor state during the RSI. The red square corresponds to regions where PES is observed, and the blue ones where PEQ is observed (the darker the color, the stronger the effect). The black dashed squares correspond to specific regions where Panels B and C zoom. (B) PES effect (ms) with respect to the coherence level at $I_{CD,max}=0.047$~nA. The light blue zone corresponds to the bootstrapped~\citep{Efron1994AnBootstrap} confidence interval at $95\%$. (C) PES effect (ms) with respect to the coherence level at $I_{CD,max}=0.035$~nA. The light blue zone corresponds to the bootstrapped confidence interval at $95\%$. }
					\label{fig9} 
				\end{figure}
					
                We studied the occurrence of the PES effect in the model with respect to the coherence level and $I_{CD,max}$, at an intermediate RSI value of $500$~ms, leading to the phase diagram in Figure~\ref{fig9}.A. We find a large domain in parameter space showing PES effect (in red in the figure). Figure~\ref{fig9}.B zooms on a  value of $I_{CD,max}$ for which PES occurs ($I_{CD,max}=0.035$~nA). We observe that the magnitude of the PES effect goes from zero to ten milliseconds at $c=10\%$, hence remaining within the range of behavioral data~\citep{Jentzsch2009,Danielmeier2011} ($10-15$~ms for a RSI of $0.5 - 1$ second). 
                In these experiments (a flanker task with stimuli belonging to one of two opposite categories, Left or Right directions), the ambiguity level is not quantified. However, the observed error rates are found around $10\%$ which, within our model, corresponds to a coherence level of about $c=10\%$.
               On the phase diagram, one can observe the variation of the PES effect with respect to the coherence level. In the region where we observe a PES effect, we find that it is enhanced under conditions when errors are infrequent. However, for large values of the coherence level, this effect cannot be observed anymore due to the absence of any error in the successive trials (almost $100\%$ of correct trials). This occurrence of PES, principally at low error rates, has been found in experiments of \cite{Notebaert2009b,NunezCastellar2010}, for which the authors observe PES when errors are infrequent, but not when errors are frequent.       
             Note that these experiments are with 4-AFC tasks, but we expect the same type of properties as for TAFC tasks -- and the model could easily be adapted to such cases with a neural pool specific to each one of the four categories.

The phase diagram, Figure~\ref{fig9}.A, also shows parameter values with no effect at all (in gray), and a domain with the opposite effect, that is with reaction times faster after an error than after a correct trial (in blue). We propose to call this effect \emph{post-error quickening} (PEQ), as opposed to post-error slowing.   As shown in Figure~\ref{fig9}.C, we find that, for a given value of $I_{CD,max}$, one can have PES at low coherence level, and PEQ at high coherence level.         

 This  PEQ effect, although much less studied, has been observed in various AFC experiments, either without feedback~\citep{Rabbitt1977,Notebaert2009b,King2010} 
 or with feedback~\citep{Purcell2016}, notably for fast-response regimes~\citep{Notebaert2009b,King2010}. The conditions for observing PEQ remain however not well established, with some contradictory results. 
We note that with Go/no-go protocols (which are similar to AFC protocols in many respects),   \cite{Hester2005} report post-error decrease in reaction times for aware errors, but not for unaware errors, whereas \cite{Cohen2009} on the contrary reports no PEQ effect, but a larger PES effect for aware errors than for unaware errors. The fact that the model predicts PEQ in TAFC tasks at high coherence levels is more in line with the results of the fMRI experiments of \cite{Hester2005}. Indeed, at high coherence levels, responses are fast and most often correct. In the rare case of an error,  the subject is likely to become aware that an error has been made~\citep{Yeung2012}.               
This thus may lead to a correlation (without causal links) between aware errors and PEQ.

\begin{figure}
					\centering
					 \includegraphics[width=0.5\textwidth]{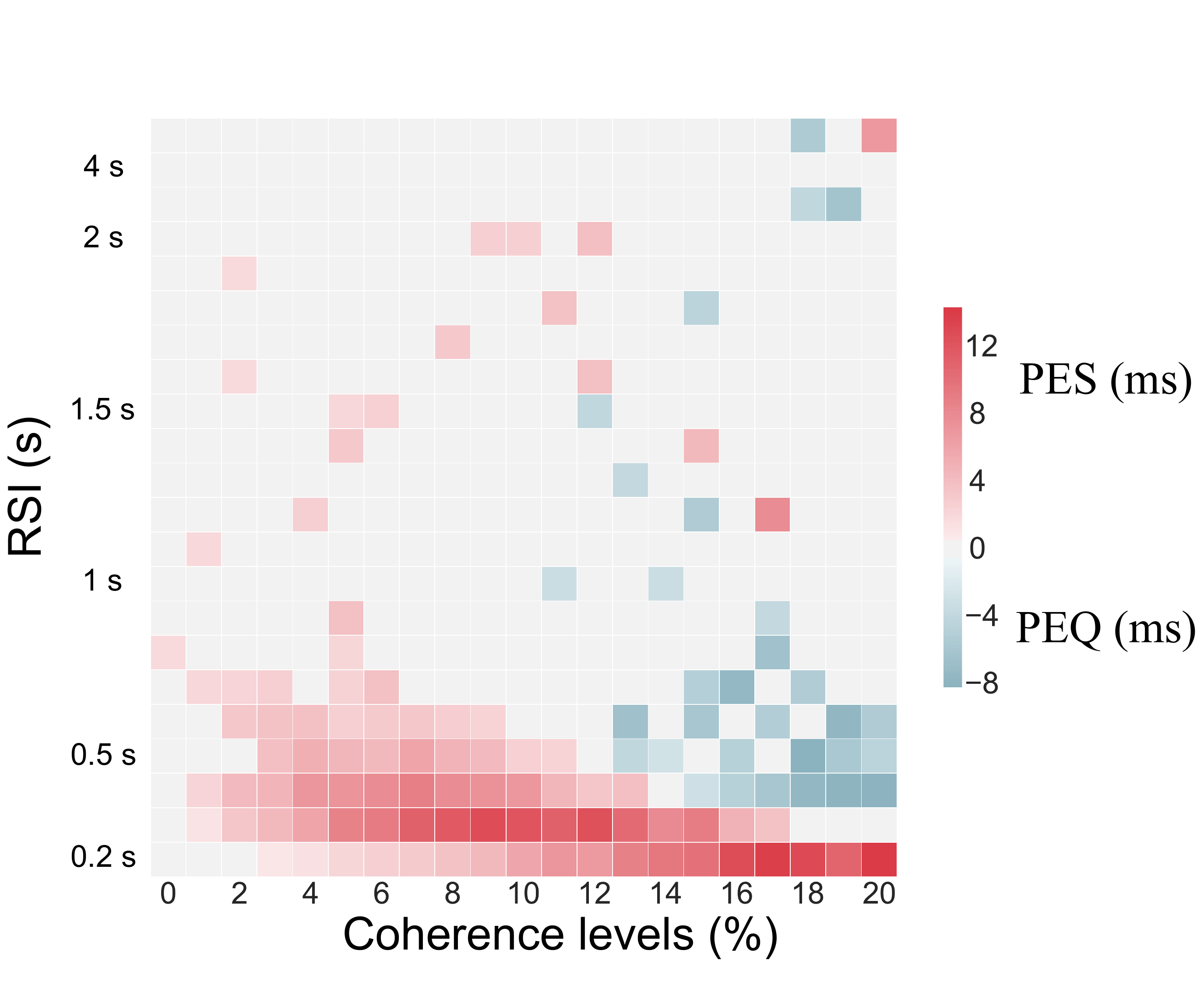} 
					\caption{{\bf Post-error slowing depending on RSI.} (A) Phase diagram of the PES effect at $I_{CD,max}=0.045$~nA. The red square corresponds to regions where PES is observed, and the blue ones where PEQ is observed (the darker the color, the stronger the effect). We used a bootstrapped confidence interval in order to decide whether or not PES/PEQ is observed.  }
					\label{fig10}
				\end{figure}

   We also studied the RSI dependency of the PES effects by plotting the phase diagram at $I_{CD,max}=0.045$~nA with respect to the RSI (Figure~\ref{fig10}). In behavioral experiments the PES effect depends strongly on the RSI. For RSI longer than $1000-1500$~ms the observation or not of PES depends specifically on the decision task \citep{Jentzsch2009,King2010}. However, a common observation is that, whenever PES is observed, if one keeps increasing the RSI, 
                the PES effect eventually disappears. 
              In Figure~\ref{fig10}, we observe that, for parameters where PES is observed at a RSI of $500$~ms,  increasing the RSI leads to the weakening of the post-error slowing effect until its disappearance. At a RSI of $1000-1500$~ms this effect is not present anymore, in agreement with experimental results~\citep{Jentzsch2009}.

The variation of PEQ with respect to RSI has not been experimentally studied, as this effect is more controversial. However, our model shows that the dependence on RSI is similar to the one of PES, and predicts that when both effects exists at a same RSI value (for different coherence levels), increasing the RSI leads to the disappearance of both of them.

We note here that the set of phase diagrams that we present in this work on the various effects, 
Figures~\ref{fig9} to \ref{fig12}, 
provide testable behavioral predictions. As just discussed in the particular case of PES and PEQ , 
they predict how the effects on reaction times are or are not correlated, and in particular how they qualitatively depend on, and co-vary with, the coherence level or the duration of the RSI.

				\subsubsection*{Post-error improvement in accuracy}
				
				Post-error improvement in accuracy (PIA) is another sequential effect reported in experiments~\citep{Laming1979,Marco-Pallares2008,Danielmeier2011}. 
                                PIA has been observed on different time-scales: long-term learning effects following error \citep{Hester2005} and trial-to-trial adjustments directly after commission of error responses. We only consider this latter type of PIA. The specific conditions under which PIA can be observed  in behavioral experiments have not been totally understood. 
               We investigate this effect in the specific context of our model, considering that the strength of the effect is linked to the difference in error rates between post-error and post-correct trials.
                
\begin{figure}
					\centering
					\includegraphics[width=0.9\textwidth]{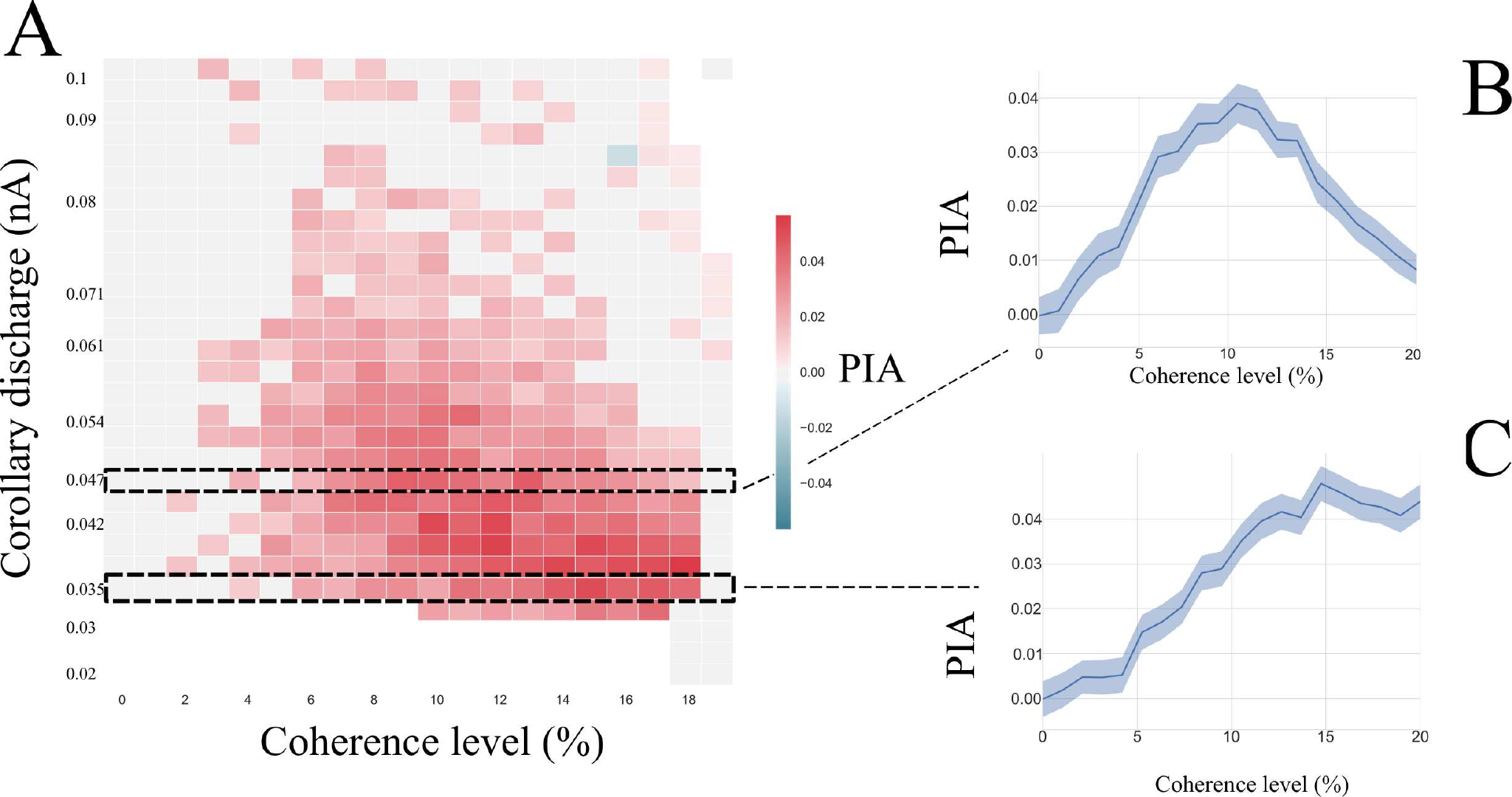}
					\caption{{\bf Post-error improvement in accuracy at a RSI of 500 ms.} (A) Phase diagram of the PIA effect at RSI of $500$~ms. The bottom white zone corresponds to parameters where sequential decision-making is impossible. The red square corresponds to regions where PIA is observed. The black dashed squares correspond to specific regions where panels B and C zoom. (B) PIA effect with respect to the coherence level at $I_{CD,max}=0.047$~nA. The light blue zone corresponds to the bootstrapped confidence interval at $95\%$. (C) PIA effect with respect to the coherence level at $I_{CD,max}=0.035$~nA. The light blue zone corresponds to the bootstrapped confidence interval at $95\%$. 
					}		
					\label{fig11}
				\end{figure}

				In Figure~\ref{fig11} we represent the phase diagram of the PIA effect with respect to coherence levels (x-axis) and corollary discharge amplitude (y-axis). We denote a large region of parameters for which PIA is present. We find a magnitude of the PIA effect of about $2-4\%$, which is of the same order of magnitude as in the experiments where, for RSIs in the range $500-1000$~ms, it is found that post-error accuracy is improved by approximatively $3\%$~\citep{Jentzsch2009}.

				Looking at Figure~\ref{fig11}, one sees that the PIA and PES effects append in the same region of parameters. However, if we zoom in on specific regions (Figure~\ref{fig11}.B and C), we can notice some differences in the variation of these effects.
				The black dashed rectangular regions correspond to the same parameters as in Figure~\ref{fig9}. We first note that PIA is also observed in these regions. However, we observe a decrease of PES at very large coherence (Figure~\ref{fig9}.B), but not of PIA (Figure~\ref{fig11}.B). Moreover the decrease of the PIA effect in Figure~\ref{fig9}.C does not occur at the same values of parameters as for the PES one.  
            It would be tempting to interpret PIA as a better accuracy resulting from taking more time for making the decision. This is not the case, since PIA does not appear uniquely in the PES region, but in the PEQ one too. In agreement with these model predictions, \cite{Danielmeier_etal2011}, in a TAFC task with color-based categories, observe that PIA can occur in the absence of PES, but that the occurrence  of PES is always associated with PIA (except for one subject among 20, results reported in Figure~1 in \cite{Danielmeier2011}).
              
\begin{figure}			
					\centering			
				\includegraphics[width=0.9\textwidth]{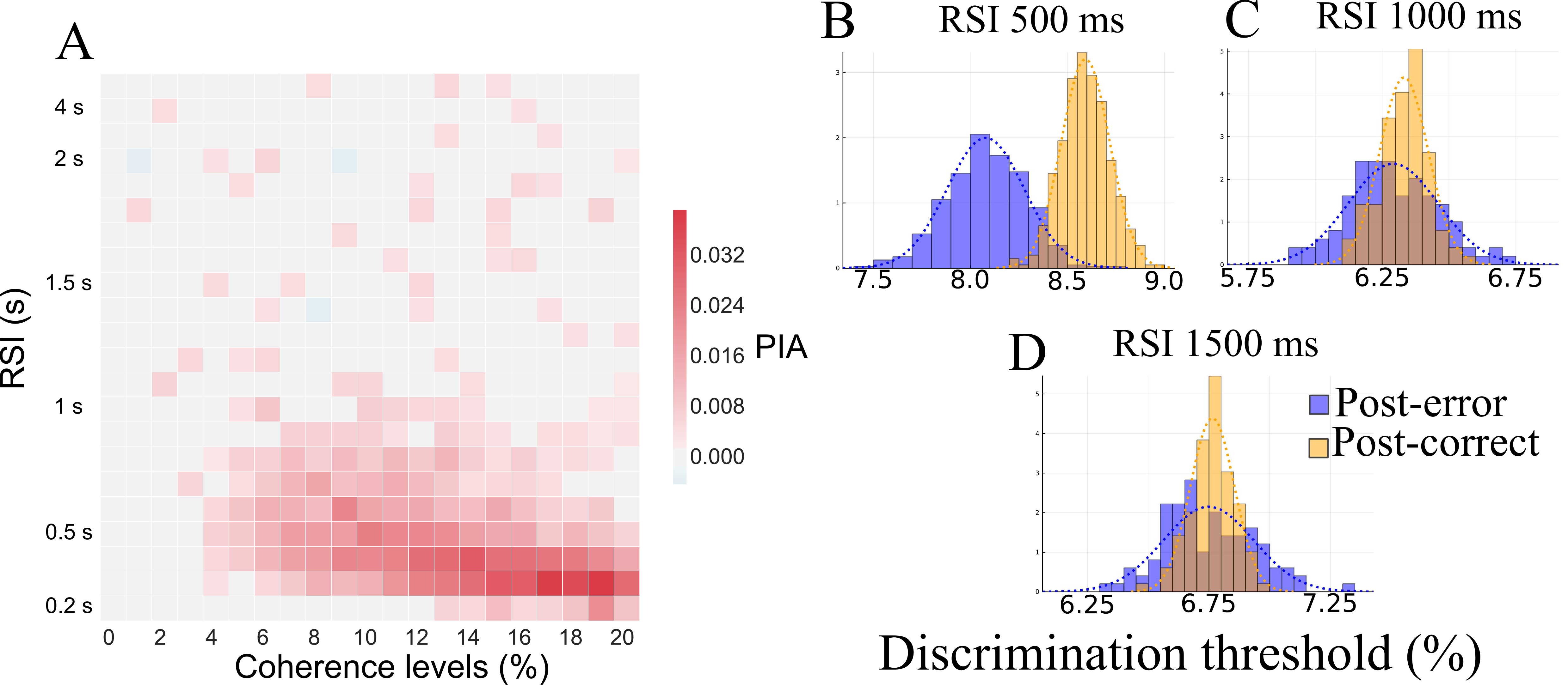} 
					\caption{{\bf Post-error improvement in accuracy depending on RSI.} (A) Phase diagram of the PIA effect at $I_{CD,max}=0.045$~nA. The red square corresponds to regions where PIA is observed. (B)-(C)-(D) Distribution of the discrimination threshold for three values of RSI ($500,1000,1500$~ms respectively). In yellow we represent the histogram of the post-correct trials, and in blue the post-error ones. The dashed curves of the corresponding color corresponds to the cumulative functions of these distributions. The corollary discharge is $I_{CD,max}=0.035$~nA.}
					\label{fig12}				
				\end{figure}

\begin{figure}[!ht]
						\centering
					\includegraphics[width=0.8\textwidth]{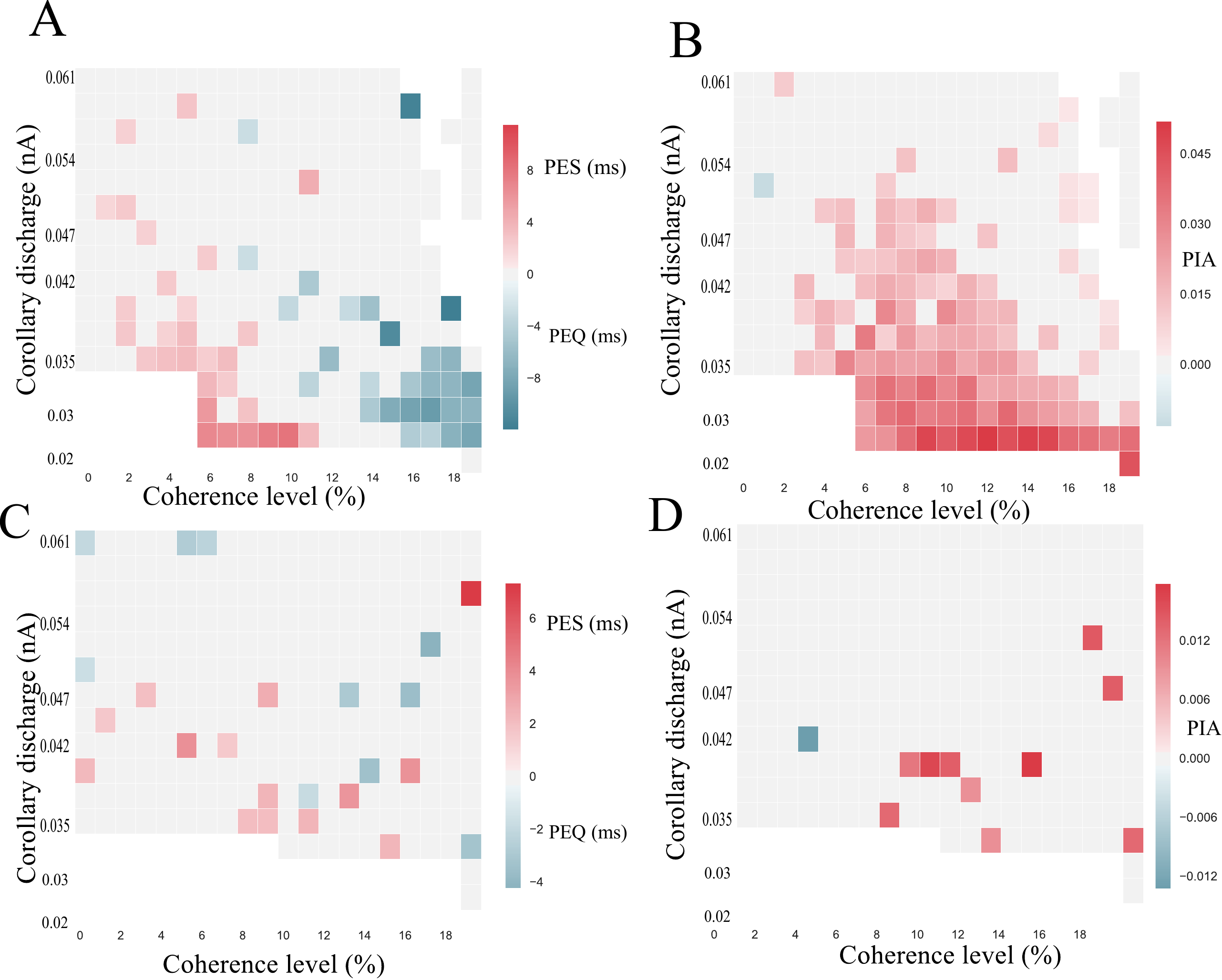} 
						\caption{{\bf Post-error adjustments at $\tau_{CD}=500$~ms (panels A and B), and second order post-error adjustments (panels C and D).} (A) Phase diagram of the PES effect. 
                        We used a bootstrapped confidence interval in order to decide whether or not PES (or PEQ) is observed. (B) Phase diagram of the PIA effect. 
                        The observation of post-error adjustments is highly impacted with the value of $\tau_{CD}$, as we do not observe PES for the same range of parameters.
                           (C) Phase diagram of the PES effect at the $n+2$ trial. (D) Phase diagram of the PIA effect at the $n+2$ trial. 
							 One sees rare isolated red squares, indicating the absence of any systematic effect. 
                             For all panels:  Simulations with a RSI of $500$~ms, other parameters as in Table~1. Color code as in Figure~9. } 
						\label{fig13}
					\end{figure}

				In EEG experiments,  \cite{Marco-Pallares2008} find that time courses of PES and PIA seem to be dissociable as 
                 they observe post-error improvements in accuracy
                 with longer inter-trial intervals (up to $2250$~ms) than post-error slowing. 
                 We note that these authors consider protocols with and without stop-signals, and here we are only concerned by those without.
                 We investigate the variation with respect to the RSI of PIA in our model (Figure~\ref{fig12}). We note that, for long RSIs, the PIA effect is not observed anymore. However as observed in \cite{Marco-Pallares2008}, the PIA effect occurs for longer RSIs than the PES effect (Figure~\ref{fig11}.A). 
				In the same way, PIA is more robust with respect to the intensity of the corollary discharge. This is corroborated by Figure~\ref{fig13}-A-B, which represents PES and PIA effect for a larger relaxation time,  $\tau_{CD}=500$~ms, hence with a stronger corollary discharge. 
				We note that all the regimes previously observed are present, for slightly different parameter ranges. This shows that the global picture illustrated by the phase diagrams, Figures~\ref{fig9}, \ref{fig11}, is 
                not specific to a narrow range of $I_{CD,max}$ and $\tau_{CD}$ values.

				 \cite{Verguts2011} find that PIA and PES seem to happen independently, suggesting that at least two post-error processes takes place in parallel. An important outcome of our analysis is to show that PIA and PES effects can both result from the same underlying dynamics. 
								In addition, in the parameters domain where they both occur, we find that the variations of these effects with respect to the coherence levels are indeed  uncorrelated (Pearson correlation test:  RSI of $500$~ms and $I_{CD}=0.035$~nA, $p=0.58$, $I_{CD}=0.05$, $p=0.79$ and $I_{CD}=0.1$~nA, $p=0.25$; RSI of $2000$~ms and $I_{CD}=0.035$~nA, $p=0.37$).
				This non-correlation highlights the complexity of such post-error adjustments, as explored in \cite{Verguts2011}.

				In order to gain more insights into the PIA effect, we study the discrimination threshold following an error or a success, with respect to the RSI (Figure~\ref{fig12}.B-D). 
In Figure~\ref{fig12}.B we represent the distribution of the discrimination threshold for $I_{CD,max}=0.035$~nA and a RSI of $500$~ms. For these parameters, the distributions for the post-error and post-success cases are highly different (Smirnov-Kolmogorov test: 
				$p<10^{-20}$). If one increases the RSI ($1000$~ms for Figure~\ref{fig12}.C and $1500$~ms for Figure~\ref{fig12}.D), this difference disappears (Smirnov-Kolmogorov test: $p=0.038$ and $p=0.4$ respectively). However, we note that the 
            model predicts a wider distribution of the discrimination threshold after an error than after a correct trial, 
                independently of the presence of the PIA effect. This might result from the wider distribution in the neural (or synaptic) activities after an error that we discuss n the next section. 
                To our knowledge, this effect has not been studied in behavioral experiments.

				\subsubsection*{Dynamical analysis}

\begin{figure}
					\centering
				\includegraphics[width=1.0\textwidth]{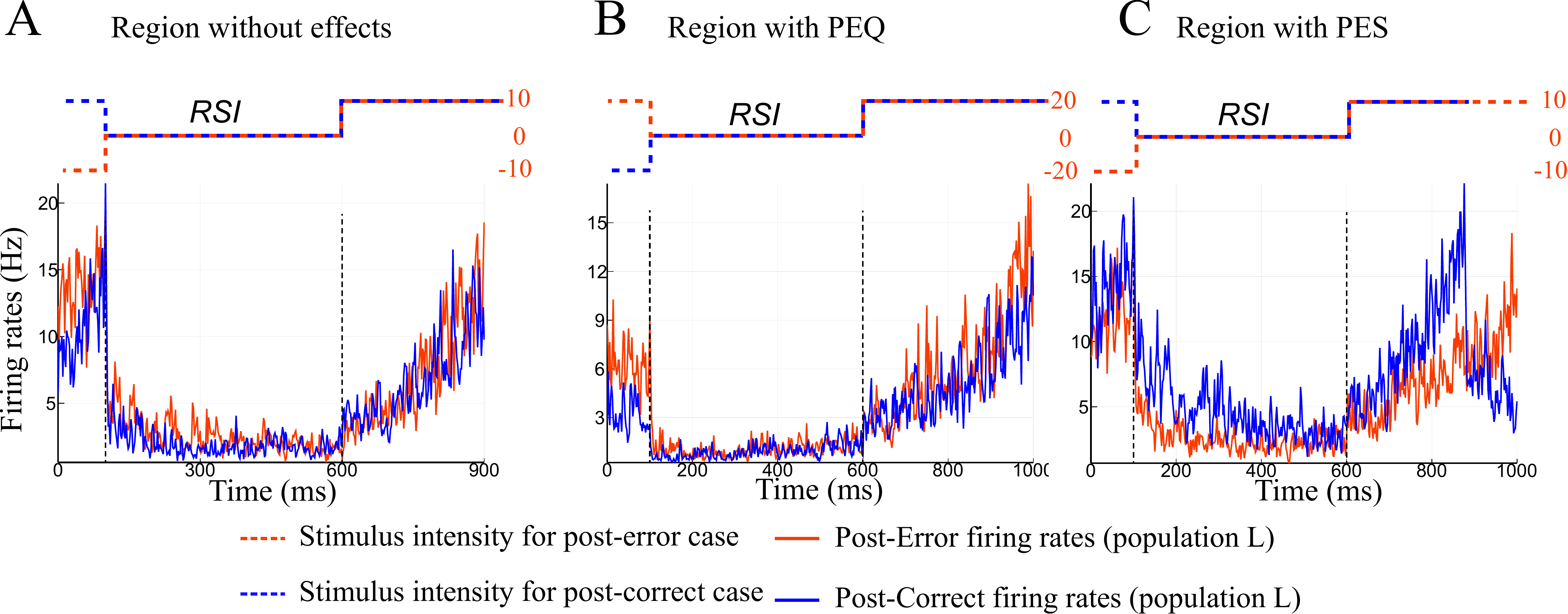} 
					\caption{{\bf Neural activities of individual trials.} (A) Dynamics for individual trials for the winning populations of the next trial: in blue the post-correct case and in red the post-error one. The dashed lines represent the coherence of the stimuli with respect to time. In blue we represent the post-correct case, and in red the post-error one. The parameters are set to a region without PES or PEQ effects ($I_{CD} = 0.047$ nA and $c=\pm 10\%$). (B) This panel represents the dynamics in the region of PEQ ($I_{CD} = 0.047$ nA and $c=\pm 20\%$). On this trial we can notice that the post-error dynamics is faster than the post-correct one. (C) The parameters are set to the PES region ($I_{CD} = 0.035$ nA and $c=\pm 10\%$). The post-correct dynamics (in blue) reaches the threshold sooner than the post-error one (in red).}
						
					\label{fig14}
				\end{figure}

				In this section we analyze the PES and PEQ effects in terms of neural dynamics.
							 First of all, we represent and discuss the dynamics on individual trials for the three regions of parameters: with neither PES nor PEQ effects, with PES effect, and with PEQ effect (Figure~\ref{fig14}). We observe the dynamics for post-error and post-correct trials during the relaxation period following a decision and during the presentation of the next stimulus. Already on individual trials we notice differences between the regions. Figure~\ref{fig14}.A represents a trial in the region without PES or PEQ. The post-error/correct dynamics are indistinguishable. Hence we do not observe any differences in the reaction times. Looking at a trial in the PEQ region (Figure~\ref{fig14}.B), we notice that the population $L$ (the winning one for the second stimulus) for the post-error case seems a bit higher in activity than for the post-correct case. This leads to the post-error quickening effect, as the post-error (orange) curve will reach the threshold sooner than the post-correct (blue) one. Finally, Figure~\ref{fig14}.C represents individual trials for parameters in the PES region. In the phase diagram (Figure~\ref{fig9}) the effect was more pronounced than PEQ, thus it is more pronounced on the dynamics too. During the relaxation, and the presentation of the next stimulus, the post-correct dynamics (blue curve) for population $L$ (the winning one for the second stimulus) is higher than the post-error one. As we can observe this leads to a faster decision time for the post-correct trial than for the post-error one. 
				
\begin{figure}[!ht]
						\centering
						\includegraphics[width=1.0\textwidth]{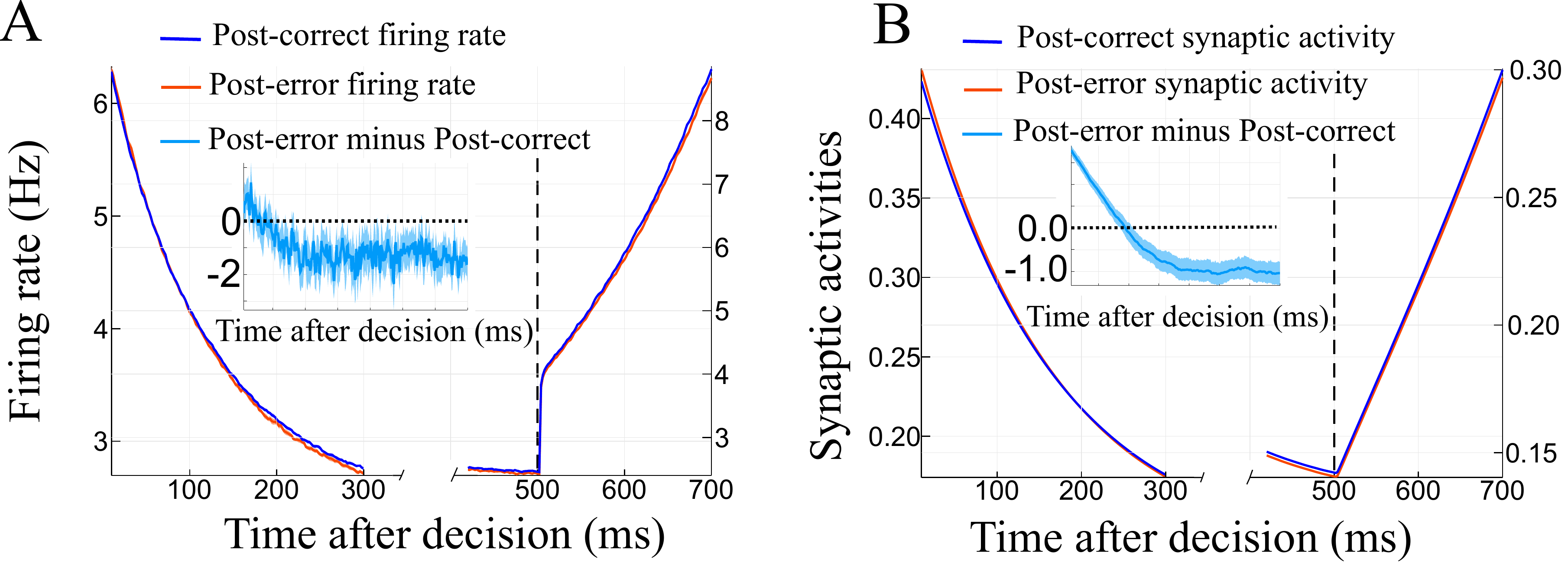} 
						\caption{ {\bf Mean firing rates of the winning population.}		 (A) Mean firing rates for the winning populations of the next trial: in blue the post-correct case and in orange the post-error one. The ribbons represent the $95\%$ confidence interval on 25 simulations(bootstrap method). The left-axis represents the relaxation of the dynamics. The right-axis is for the beginning of the next stimulus. The parameters are set to a region with PES effects ($I_{CD} = 0.035$ nA and $c=\pm 10$). The sub-panel with the light blue curve is the difference between the post-error firing rates and the post-correct with respect to time (in percent). The ribbon stands for the $95\%$ confidence interval. As expected, this difference is negative. Hence the post-correct dynamics is faster and crosses sooner the threshold. This leads to the PES effect. 
                        (B) Mean synaptic activities for the winning populations of the next trial: in blue the post-correct case and in orange the post-error one. The sub-panel with the light blue curve is the difference between the post-error synaptic activities and the post-correct with respect to time (in percent). 
                        }				
						\label{fig15}
					\end{figure}

\begin{figure}
					\centering
				 \includegraphics[width=1.0\textwidth]{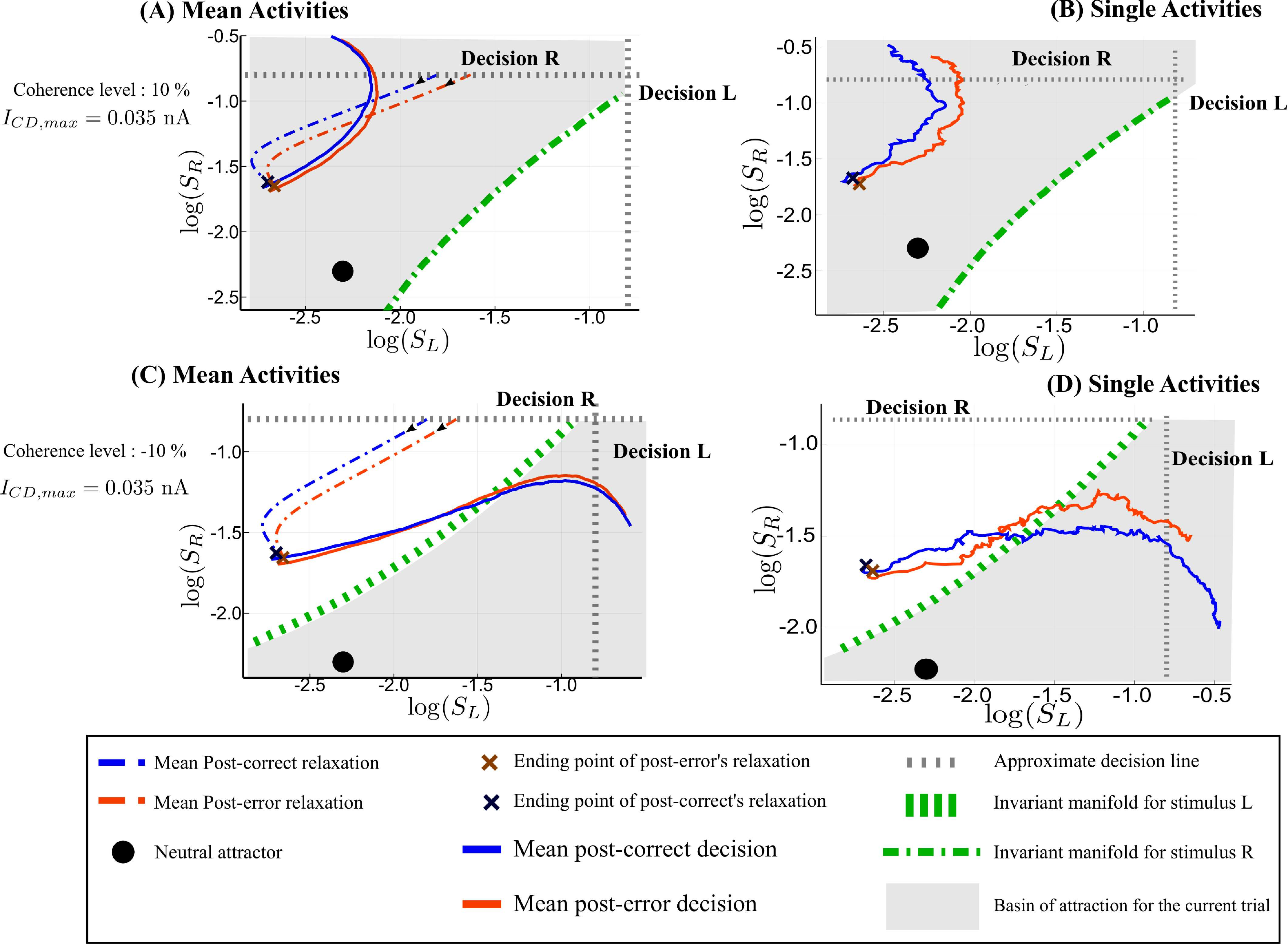}  
					\caption{ {\bf Analysis of the post-error trajectories for the PES regime.} Phase-plane trajectories (in log-log plot, for ease of viewing) of the post-correct and post-error trials. We consider that the previous decision was decision R. The black filled circle shows the neutral attractor state (during the relaxation period). During the presentation of the next stimulus, the attractors and basins of attraction change (represented by the gray area and the green dashed lines). Panels (A) and (B): PES and PIA regime ($c=10\%$ and $I_{CD,max}=0.035$~nA) in the repeated case. The blue color codes for post-correct trials, and the red one for post-error. Panel (A): average dynamics; Panel (B): single trajectories during the next trial.
							Panels (C) and (D): regime with PES and PIA in the alternated case ($c=-10\%$ and $I_{CD,max}=0.035$~nA). 
							The dynamics after the relaxation is followed during $400$ms for repeated and $800$~ms for alternated case, as if there were no decision threshold. The actual decision occurs at the crossing of the dashed gray line, indicating the threshold. 
					} 
					\label{fig16}
				\end{figure}

\begin{figure}
					\centering
					\includegraphics[width=1.0\textwidth]{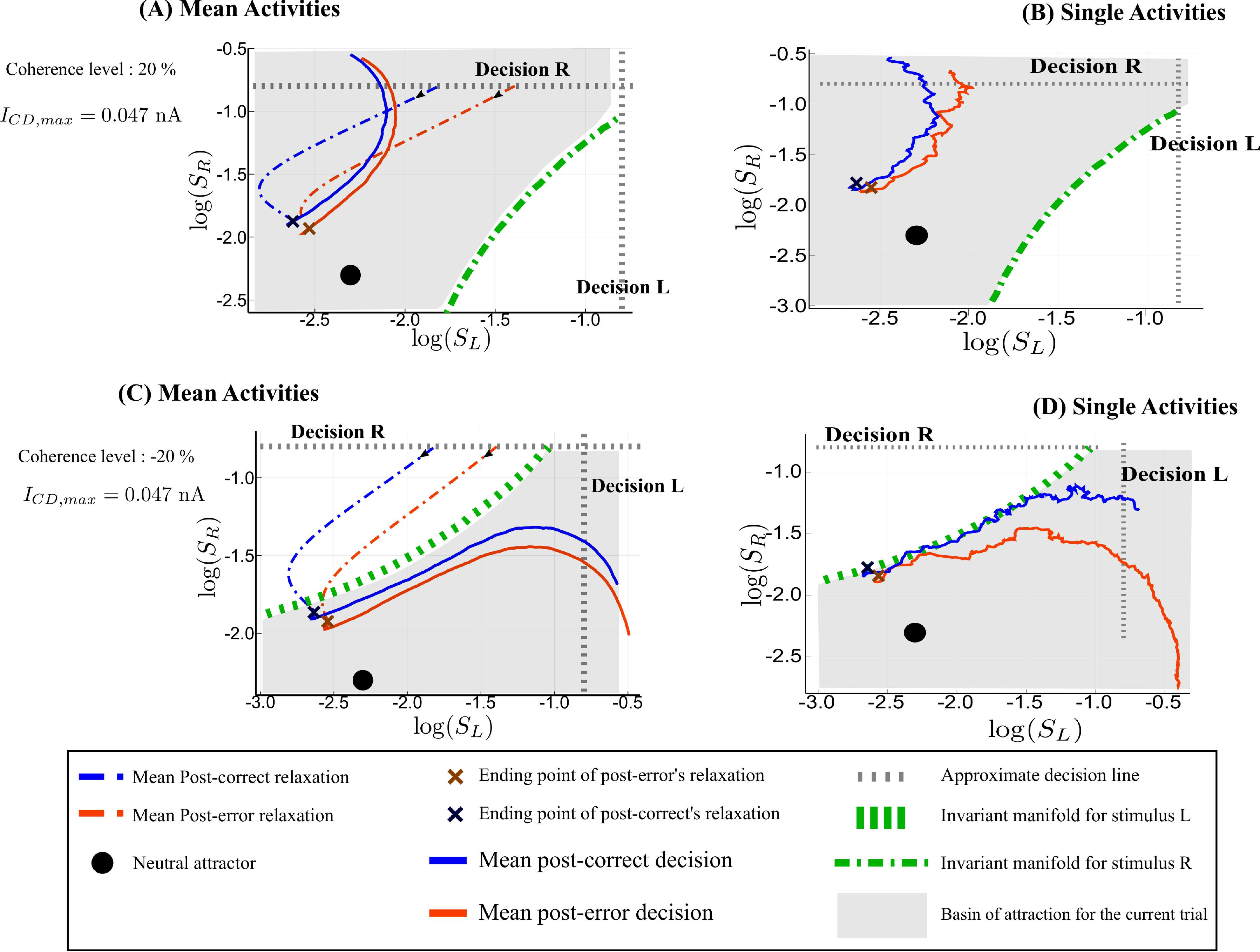} 
					\caption{{\bf Analysis of the post-error trajectories in the PEQ regime.} Phase-plane trajectories (in log-log plot, for ease of viewing) of the post-correct and post-error trials (same as Figure~\ref{fig16} in the PEQ case). We consider that the previous decision was $R$. The black filled circle shows the neutral attractor state (during the relaxation period). During the presentation of the next stimulus, the attractors and basins of attraction change (represented by the gray area and the green dashed line). Panels (A) and (B): PEQ and PIA regime ($c=20\%$ and $I_{CD,max}=0.047$~nA). The blue color codes for post-correct trials, and the red one for post-error. The plain lines represent mean dynamics for  (A) or single dynamics (B). 
							Panels (C) and (D): regime with PEQ and PIA in the alternated case ($c=-20\%$ and $I_{CD,max}=0.047$~nA).The post-error relaxation already lies within the alternated basin of attraction. For alternated trials, the dynamics needs to cross the invariant manifold (green dashed line), which denotes the boundary between the basins of attraction.   The dynamics is followed during $400$ms for repeated and $800$~ms for alternated case, as if there were no decision threshold. The actual decision occurs at the crossing of the dashed gray line, indicating the threshold. 
					} 
					\label{fig17}
				\end{figure}

\begin{figure}[!ht]
						\centering
						\includegraphics[width=1.0\textwidth]{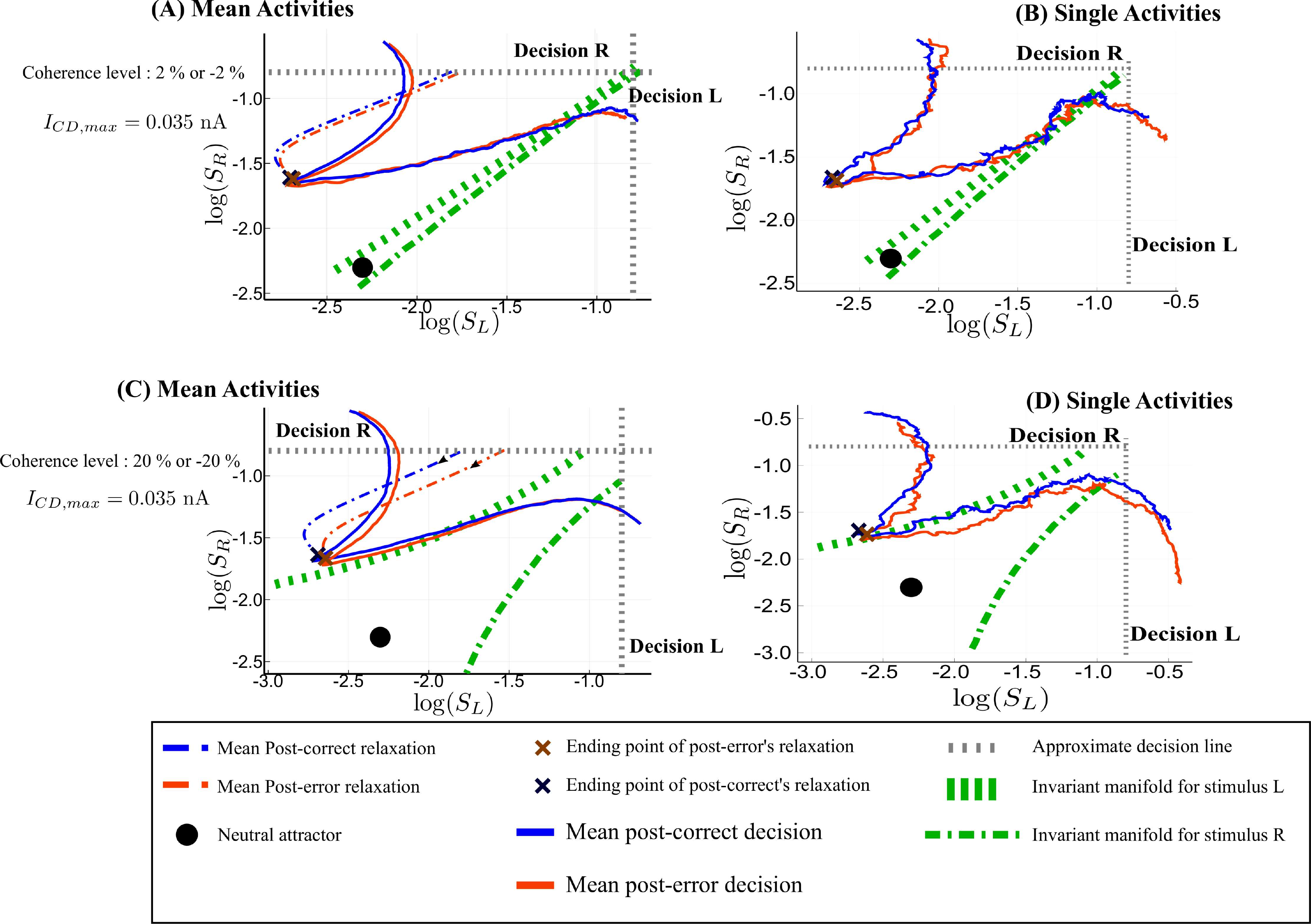}
						\caption{ {\bf Analysis of the post-error trajectories in the regime with neither PES nor PEQ effect} Phase-plane trajectories (in log-log plot, for ease of viewing) of the post-correct and post-error trials. We consider that the previous decision was decision R. The black filled circle shows the neutral attractor state (during the relaxation period). During the presentation of the next stimulus, the attractors and basins of attraction change (represented by the gray area and the green dashed lines).
							Panels (A) (mean dynamics) and (B) (single dynamics): regime without PES or PIA ($c=\pm 2\%$ and $I_{CD,max}=0.035$~nA). We show both the alternated and the repeated case, with the corresponding basins of attraction. The blue color codes for post-correct trials, and the red one for post-error. For alternated trials, the dynamics needs to cross the invariant manifold (green dashed line), which denotes the boundary between the basins of attraction.
							Panels (C) (mean dynamics) and (D) (single dynamics): regime with PIA but without PES ($c=\pm 20\%$ and $I_{CD,max}=0.035$~nA).   The dynamics is followed during $400$ms for repeated and $800$~ms for alternated case, as if there were no decision threshold. The actual decision occurs at the crossing of the dashed gray line, indicating the threshold.
						}  
						\label{fig18}
					\end{figure}

					We show now that the dynamics explains the three effects PES, PEQ and PIA. 
					We provide in Figures~\ref{fig16},~\ref{fig17} and~\ref{fig18} a semi qualitative and semi quantitative analysis of the dynamics of the synaptic activities in the phase plane of the system, for several parameters regimes. Here again, the analysis is easier working on the synaptic activities. This can be seen by considering Figure~\ref{fig15} on which we represent the mean firing rate and synaptic activity of the winning population in the PES case. 
               Due to the range of variation of the firing rates, and the intrinsic noise of the system, it is hard to observe a difference between the neural activities. However, if we compute this difference (sub-panel of Figure~\ref{fig15}) we note the following.
                    At the beginning of the next trial, the difference between the post-error and post-correct firing rates is significantly below zero, hence the reaction time will be shorter for post-correct than for post-error trials. 
                   We  find  the same behavior for the synaptic activities~(Figure~\ref{fig15}.B), but  much less noisy, as expected from the discussion in the  Material and Methods section.

					\paragraph{PES effect.}
					We now detail the analysis of the PES effect (and of the concomitant PIA effect) based on Figure~\ref{fig16}. 
				Let us first explain how each panel is done. Without loss of generality, we assume that the last decision made is $R$. Repeated and Alternated cases thus correspond to next trial decisions $R$ and $L$, respectively. The $x$ and $y$ axis are the synaptic activities $S_L$ and $S_R$, respectively -- hence, the losing and wining populations for the first trial. 
				
					On the left panels, we represent with dashed lines the average dynamics during the relaxation period, that is from the  decision time for the previous stimulus to the onset of the next stimulus. This allows to identify clearly the typical neural states at the end of the relaxation period. 	
					The average is done over post-error (resp. post-correct) trajectories  sharing a same state at the time of the last decision. The choice of these two initial states is based on the following remark. A typical trial with a correct decision will lead, at the time of decision,  to losing and wining populations with highly different activity rates, hence a 	neural activity, and thus a synaptic activity $S_L$, 	far from the threshold value. On the contrary, a typical error trial will show a losing activity not far from the threshold -- this can also be observed in Figure 4B in \cite{Wong2007}. We can thus represent post-correct trials, respectively post-error trials, by dynamics with initial states having a rather small, respectively large, value of $S_L$ (and in both cases the first trial winning population $S_R$ at threshold value).\\
					We then represent with a continuous line the average trajectory following the onset of the next stimulus. We  observe this dynamics during the same time for post-error and post-correct cases, as if there were no decision threshold, in order to compare the dynamics of post-error and post-correct cases for the same duration of time. 
				Decision  actually occurs when the trajectory crosses the decision line (dashed gray line) -- this is approximate: because of the noise, there is no one to one correspondence between a neural activity reaching the decision threshold and a particular value of the associated synaptic activity. Having all the trajectories plotted for the same duration (and not only until the decision time) allows to visually compare the associated reaction times.
				
				On the right panels, we represent typical trajectories during the presentation of the next stimulus.		The black dot on every panel gives the location of the neutral attractor that exists during the relaxation dynamics. The basins of attractions that are represented are the one associated with the attractors $L$, $R$, of the dynamics induced by the onset of the next stimulus. Remind that these attractors are different from the ones associated to the dynamics during the relaxation period.
				
				We can now analyze the dynamics. In the repeated case (Figure~\ref{fig16}.A and B), at the end of the relaxation (that is at the onset of the next stimulus), both post-correct and post-error trials lie into the correct basin of attraction. Hence, the error rates for these trials are similar. However, the neural states reached at the end of the relaxations are different. Compared to the  post-error trial, the post-correct state is closer to the boundary of the new attractor associated to decision $R$, and the corresponding decision will thus be faster. In the alternated case (Figure~\ref{fig16}.C and D), the states reached at the end of the relaxation period do not lie within the correct basin of attraction. During the decision-making dynamics, the trajectory needs to cross the boundary between the two basins of attraction. 
				The post-correct trials leading to an alternate decision have a rather straight dynamics across the boundary, leading to relatively fast decision times. In contrast, the states at the onset of the stimulus  of the post-error trials are closer to the boundary so that the corresponding trajectories crosses with a smaller angle with respect to the basin boundary. This leads to longer reaction times, hence the PES effect. 
                It would be interesting to have electro-physiological data with which the model predictions could be directly compared. 
                However, in a typical experiment on monkeys, a feedback on the correctness of the decision is given, since the animal learns the task thanks to a reward-based protocol.
               Nevertheless, we note that, in the random dot experiments on monkeys of \cite{Purcell2016}, the authors find a higher buildup rate of the neural activity for post-correct trials than for post-error trials (see Figure 6 in  \cite{Purcell2016}). Within our framework, this can be understood as trajectories that cross the basin boundary more quickly for post-correct trials, in accordance with our model's predictions. 
                This suggests that the observed difference in buildup rates may not result from some mechanism making use of the information on the correctness of the decision, but rather from the nonlinear dynamics discussed here.

				The PIA is understood from the same analysis as for the PES effect. For specific realizations of the noise that lead to error trials, the post-error trials dynamics is closer to the boundary. Thus it has a higher probability to fall on the other side of the basin of attraction. Hence, the error rates are lower for post-error trials than post-correct trials.
				
					\paragraph{PEQ effect.}
					The PEQ effect can be understood from the same kind of analysis, based here on Figure~\ref{fig17} (analogous for the PEQ effect to Figure~\ref{fig16}  for the PES effect).   As seen previously, the PEQ effect occurs mostly at high level of coherence.
				We consider first the repeated case (Figure~\ref{fig17}.A and B). Since the coherence level is high, at the end of the relaxation period, both post-correct and post-error trials lie within the correct basin of attraction, far from the basin boundary. The reaction times and error rates of post-correct and post-error trials for repeated decisions are thus similar. 
				
				In contrast, the alternated case (Figure~\ref{fig17}.C and D) exhibits both the PIA and the PEQ effects. The post-error's end of relaxation now is inside the basin of attraction of the alternated choice. 
				Hence, the error rate will be lower than when the ending point is outside this region (post-correct trials begin at the boundary of the basin of attraction). Moreover, the post-correct trials dynamics have to cross the boundary.   Hence they are closer to the manifold, which lead to slower dynamics, whereas the post-error dynamics can directly reach the new attractor state.
				This analysis explains why the decreasing of PES and PIA do not occur at the same coherence level too. Indeed the decreasing of PIA occurs when the ending point of the post-error relaxation crosses the boundary, whereas the post-correct ending point remains into the same basin of attraction. For the PES effect to decrease, the dynamics for both cases just need to be closer to the boundary and not necessarily on the opposite side. Hence the decrease of the PES effect occurs at lower coherence than the PIA one. \\
                 Here we have seen that the occurrence of the PEQ effect depends on some very specific and fragile feature, the crossing or not of a basin boundary. The conditions for observing the effect are thus likely to vary from individual to individual, and from experiment to experiment. This may explain why the experimental results about the PEQ effect remain controversial.

				In Figure~\ref{fig18} A and B we investigate the parameter regime, at low coherence level, for which there is no effect -- neither PES, nor PEQ or PIA. The post-error and post-correct dynamics are very similar and lead to the same relaxation ending point, far from the basin boundary.     
				Finally, in Figure~\ref{fig18} C and D we consider the parameter regime, at high coherence level, with only the PIA effect. Here the relaxations of post-error and post-correct trials are different. However, as for the PEQ effect, at high coherence level both dynamics will be fast. For alternated trials, none of the two ending points are in the correct basin of attraction. 
				
            As discussed for the PES effect, electro-physiological data only exist for experiments with feedback on the correctness of the decision. In experiments on monkeys, \cite{Purcell2016} obtain puzzling results for what concern the PEQ effect. They observe an important difference in baseline activities for post-correct and post-error trials, which is not well accounted for either by their DDM analysis or by our model. 
              However, in terms of neural dynamics, this observed difference in the level of neural activities obviously implies that the dynamical states are different at the time of the onset of the stimulus, a fact in agreement with our model's predictions. 
              One may wonder if the separation in baseline activities, and not just in starting points, could be a consequence of the feedback.

					\paragraph{Correlating post-error effects with the activity distributions at the previous decision.} To go beyond the above analysis on the post-error adjustments 
                    (PES, PEQ and PIA effects), we analyze the respective influence of the winning and losing population levels of activity at the time of the previous decision, onto the decision at the next trial. This will first confirm the previous analysis, but also provide more insights on the the specificity of the two opposite effects, PES and PEQ.

              The mean activity, at the time of the decision, of the winning population is indistinguishable between post-correct and post-error trials (Unequal Variance (Welch) test: Fail to reject, $p=0.16$ at RSI of $500$ms and  Fail to reject, $p=0.87$ at RSI of $2000$ms). However, for short RSIs (corresponding to PES regime) the mean synaptic activities, at the time of the decision, of the losing population are different for post-correct and post-error trials are different for post-correct and post-error trials (Unequal Variance (Welch) test: Reject, $p=2.7 \times 10^{-20}$  at RSI of $500$ms and  Fail to reject, $p=0.57$ at RSI of $2000$ms).

\begin{figure}
					\centering
		\includegraphics[width=1.0\textwidth]{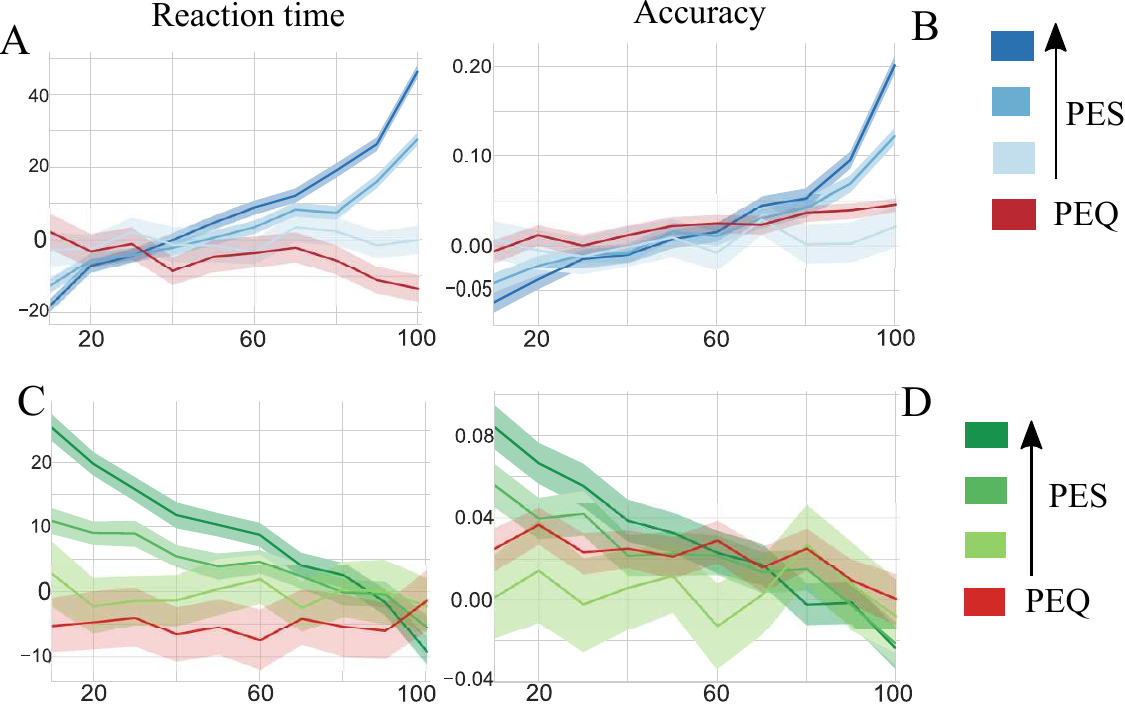}
					\caption{{\bf Influence of the losing and winning population on the post-error adjustments.}  Panels (A) and (B) represents respectively the reaction time (PES effect) and accuracy (PIA effect) with respect to inter-percentiles range of the losing population synaptic activity distribution, at a RSI of $500$~ms. The red curve corresponds to $c=18$ and $I_{CD,max}=0.047$, where we observe PEQ. Dark blue corresponds to strong PES effect ($c=10$, $I_{CD,max}=0.035$), medium blue to medium PES effect ($c=5$, $I_{CD,max}=0.05$). Light blue corresponds to no effect at all ($c=10$, $I_{CD,max}=0.035$), for a RSI of $2$ seconds.
						Panels (C) and (D) represent the same curves for the winning population, with the same color code. The shadow area represents the $95 \%$ bootstrapped confidence intervals of the corresponding effect.
					}
					\label{fig19}
				\end{figure}

				To get more insights, we plot  in Figure~\ref{fig19} the amplitude of the PES effect with respect to the inter-percentile range of the distribution of the synaptic activities of the winning and losing populations at the time of the previous decision.
				We note that when PES occurs, the higher the activity of the losing population at the time of decision, the stronger this effect will be. The influence of the winning population is observed, although in an opposite way. 
				When PES occurs these effects are correlated (Dark Blue: Pearson correlation: $r=-0.98$ and $p=2.6 \times 10^{-7}$, Medium Blue: $r=-0.98$ and $p=9.5 \times 10^{-7}$), in the sense that the variations with respect to the inter-percentile of the winning and losing population are correlated. 
				These observations are consistent with the analysis of the PES phase-plane trajectories. Indeed, the higher the losing population activity is, the closer to the invariant manifold the state at the end of the relaxation period will be. Hence, the effect will be stronger as it becomes easier (more likely) to cross the boundary.\\

				However, we observe in Figure\ref{fig19}, panels A and C, a different behavior for the PEQ effect: there is an almost constant value of the PEQ effect with respect to the inter-percentiles of the distributions of the winning and losing populations activities.  This is explained by the fact that, at the end of the relaxation, if the category of the next stimulus is the opposite of the  previous decision, the network state finds itself within the (correct) associated basin of attraction, but very close to the boundary. This is true whatever the correctness of the previous decision. However, the post-correct case will lead to an even closer location from the basin boundary. The nonlinearity of the dynamics near the basin boundary will strongly amplify the small difference between post-correct and post-error ending point. The PEQ effect will thus not be correlated with the size of this difference.

				For what concerns the PIA effect, we observe in Figure~\ref{fig19}.C-D a similar dependency in the synaptic activities as for the PES effect, with a stronger effect for high activities of the losing population. This corroborates the above phase plane analysis of the trajectories (Figure~\ref{fig13}). Indeed, the PES and PIA effects both depend on the position of the relaxation in the phase plane. Being closer to the boundary (high activity of the losing population) leads to a smaller error rate in the next trial.

From the above analysis, a  prediction of the model is that, whenever there are PES or PIA effects, the mean activity of the {\em losing} population is  different for correct and error decisions. Moreover, this level of activity
is correlated with the amplitude of the post-error adjustment effect. This can be seen in Figure~\ref{fig19}, panels A, B. In this figure we present the quantiles of the synaptic activities. The results would be similar, but much more noisy, for the firing rates.  We expect that this prediction can be tested in experiments by measuring the correlation between the amplitude of the PES (or PIA) effect, and the difference in mean activities of the losing neural population (difference between post-error and post-correct trials).

				\section*{Discussion}

				 We have shown that,  without fine tuning of parameters, an attractor neural network accounts, qualitatively and with the correct orders of magnitude, for sequential effects and post-error adjustments reported in TAFC experiments in the absence of 
                 feedback about the correctness of the decision. 
					
\begin{figure}
					\centering
					\includegraphics[width=0.5\textwidth]{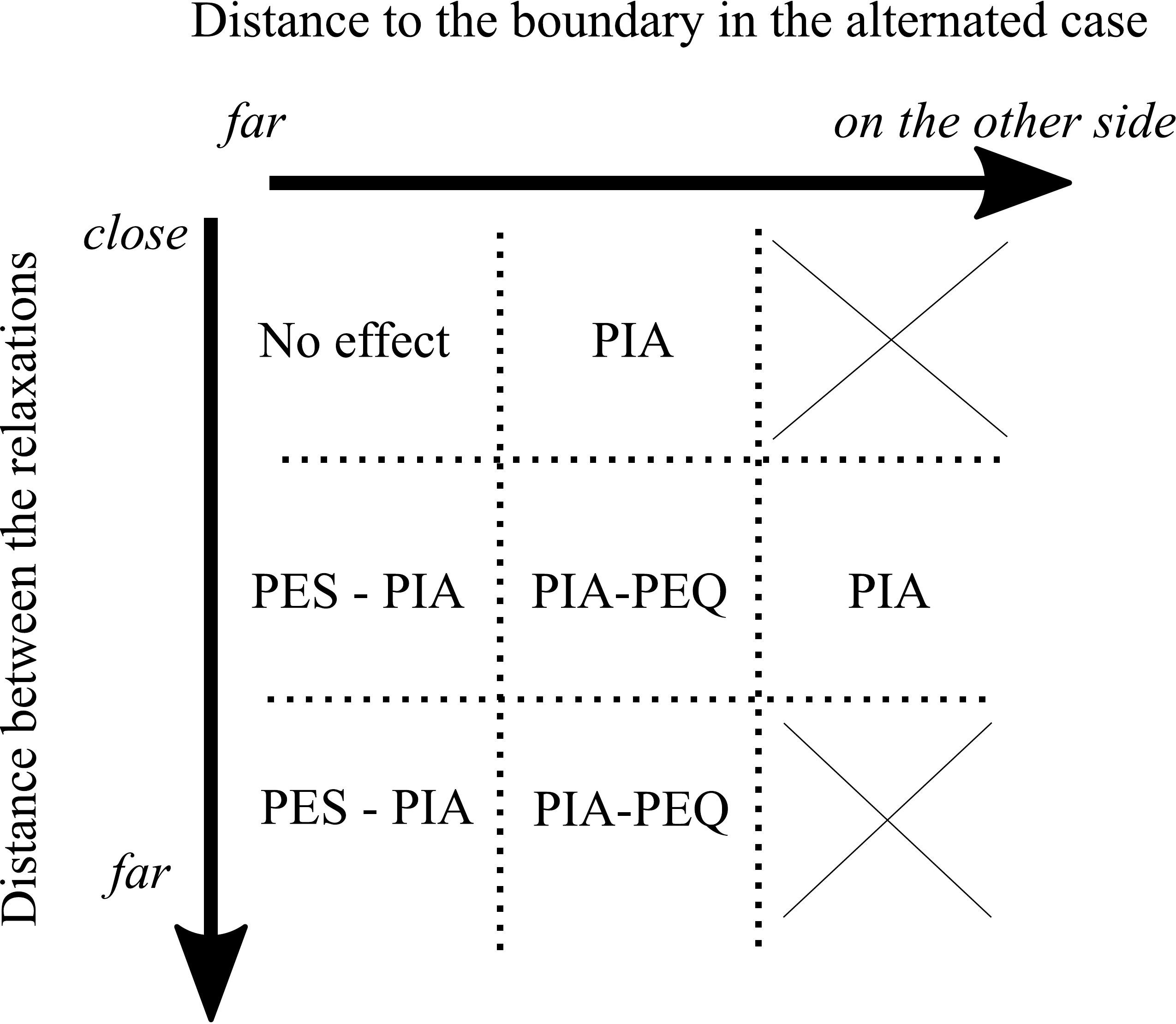}
					\caption{{\bf Schematic diagram of the post-error adjustments observations. } The {\it x-axis} represents the distance between the ending state of the relaxations and the boundary of the following basins of attraction. It goes from "the ending states are far away from the boundary" to "both ending states are in correct basin of attraction". The {y-axis} corresponds to the distance between the post-error and post-correct relaxations. The crosses denote regions which are not relevant, or inside which the network do not commit errors.}
					\label{fig20}
				\end{figure}

					We provide evidence that these effects all result from the same intrinsic properties of the nonlinear neural                  dynamics. 
				We present in Figure~\ref{fig20} a schematic diagram of the occurrence of the effects depending on the parameters, even though this does not exhaust the richness of the system’s behavior as discussed in this paper. 
				Our results suggest to test experimentally this general picture, and more precisely what is predicted by the phase diagrams, Figures~\ref{fig9}-~\ref{fig12}. In particular it would be interesting to test the occurrence of post-error quickening at large coherence level or the  variations of post-error adjustments with respect to coherence levels.

				\subsection*{Explanations for post-error slowing}

				Several cognitive explanations of PES effects have been proposed~\citep{Rabbitt1977,Laming1979a,Notebaert2009b}.

		 	In particular, these effects have been discussed in the framework of drift diffusion models~\citep{Dutilh2011,Goldfarb2012,Purcell2016}. 		
 \cite{Dutilh2011}, in experiments without feedback about the correctness of the decision, and      
     \cite{Purcell2016}, but in experiments with feedback, show that post-error and post-correct trials can be fitted  by DDMs with different sets of parameters values for post-error and post-correct trials. In addition, \cite{Dutilh2011} argue that the 
modification of the  decision threshold within the DDM framework, would correspond to the hypothesis of increased response caution, the decision becoming more cautious after an error.  Yet, the neural correlates, which would determine the threshold or the starting point remain  obscure, especially in the absence of feedback on the correctness of the trial.

				Within the attractor network framework considered here, the PES  and PEQ effects are explained 
                thanks to an in-depth analysis of the neural dynamics.
               We have shown that the location of the dynamical state at the end of the relaxation period (end of the RSI),     
                with respect to the basins of attraction of the attractors induced by the next  stimulus, depends 
                on what occurred at the previous trial.  The fact that we have different properties, e.g. for post-correct and post-error trials, for a {\em same} set of parameters values, is a result of the nonlinear dynamics which amplifies the difference in ending points of the relaxation.          
     This cannot be obtained within the DDM framework (without the addition of other mechanisms) since, in a DDM, the state reached at the time of a decision is  identical for an error and a correct trial.   
                An  additional outcome of the analysis is that, for a given set of parameters values, different regimes (PES,  PEQ or no effect) may be observed depending on the coherence level of the stimulus: due to the non linearities,
                the dynamical state   at the end of the RSI also depends on the coherence level.     
 
Typical experiment on monkeys make use of reward-based protocols, hence with feedback. This  makes difficult to have electro-physiological data in the absence of feedback. Yet, as discussed in this paper, the  faster buildup of neural activity in post-correct trials than in post-error trials, as observed by \cite{Purcell2016} on monkeys in random dot experiments, can be understood within our framework as a faster dynamics  near the boundary between attraction basins in the post-correct case. 
    
    As discussed above, another prediction of the model is that, in the case of PES or PIA, the mean activity of the {\em losing} neural population is  different for correct and error decisions, a difference which should correlate with the amplitude of the effect.

				\subsection*{First and higher order sequential effects}    
				
				Sequential effects can be categorized as first order (if 
				caused by the immediately previous trial), or higher order (if caused by earlier trials in the sequence)   ~\citep{Laming1979,Soetens1984,Soetens1985,Cho2002}. 
				Post-error adjustments have also been experimentally observed at higher order (see \cite{Laming1979}).

				Within the framework of attractor networks, the sequential effects in choice repetitions are explained by a starting bias, as discussed in~\cite{Gao2009,Bonaiuto2016} and in the present paper.    
					As stated by~\cite{Gao2009}, without any additional memory module, an attractor network cannot reproduce the transition between automatic facilitation and strategic expectancy~\citep{Laming1968}. In our network, for too short RSIs (a few dozens of milliseconds) the sequential effects are too strong to be plausible. Decision conflict mechanisms~\citep{Jones2002} could be implemented to correct this effect and to investigate other effects of repetitions and alternations~\citep{Gao2009}.

				To account for higher order effects, \cite{Gao2009} considered a dynamical network 
                 making use of additional memory modules. This network is  explicitly set up 
                 in order to 
                 reproduce automatic facilitation and strategic expectancy effects. 
				In this model, even the first order effects result from a coupling between a short-term memory module and the decision network. In contrast, we have shown here that a single attractor network, without memory units, presents first order effects as an intrinsic property of the dynamics.
				
				However, due to the nature of the dynamics in our model, we do not expect to reproduce higher-order effects. Indeed, for parameters for which the model exhibits first order sequential effects ( $I_{CD,max}=0.035$~nA),  we find neither second order sequential effects, 
                nor post-error adjustments, as illustrated in Figure~\ref{fig13}-C-D.

				One may ask whether a more complex architecture, taking into account other brain areas, could account for higher order repetition biases and post-error adjustments effects as resulting from some intrinsic properties of the dynamics,
                in the absence of specific memory units.

				\subsection*{Working memory and Decision-Making}    
				
				In this work we have considered {\it free response time task}~\citep{Roitman2002} in which the subject must make a decision as soon as possible. In the different protocol  {\it delayed visual motion discrimination experiment}~\citep{Shadlen2001}, the subject must  make the decision at a prescribed time after the onset of the stimulus. In such task, the decision choice must be stored in order to be retrieved at the prescribed instant of time.   In the original attractor neural network model (\cite{wang2002probabilistic}), the decision is stored as in a  working memory. As discussed at the beginning of this paper, within the framework of a single module of attractor decision network, the 
                corollary discharge considered in the present paper 
                allows the system to make successive decisions, at the price of removing the working memory behavior. An important issue is to understand how the decision making system can adapt itself to these  opposite contexts 
               (see \cite{niyogi2013dynamic} for a model with gain modulation). 
  It is not unrealistic to expect a control mechanism onto the inhibitory current. Depending on the task, the inhibitory current could be sent either just after the decision has been made, or later after the end of the delay period. In the latter case, a prediction is that, compared to cases without delay, there should be weaker post-error effects, but stronger repeated/alternated effects.
					
					An alternative 
                    is to have a more complex architecture.However, the memory units in  \cite{Gao2009} are not appropriate for dealing with delayed  discrimination experiments. For experiments with delays, \cite{Murray2017} consider two interacting modules, one implementing the posterior parietal cortex and an other one the posterior frontal cortex. 
It will be interesting to extend the present work by adding a working memory module in line with~\cite{Murray2017}, in order to obtain a network performing sequential decision-making while keeping the working memory behavior.
		
				 Finally we note that various brain areas have been shown to be involved in sequential decision tasks in which the memory of the last decision has to be maintained~\citep{Middlebrooks2012,Donahue2013,Abzug2018}. This suggests more generally that a broader network is necessary for decision tasks requiring  memory.

				\subsection*{Future Prospects} 
				
				During behavioral tasks, subjects are not always aware of their mistakes~\citep{Yeung2012}, but do show post-error slowing. One may thus ask why one does not generally become aware that an error has been made, since the neural dynamics is different following an error or a success. As discussed in the present work, these differences in the dynamics are very subtle. The post-error and post-correct firing rates have broad distributions, with some common properties (the same mean for example). The strong overlapping of these distributions makes it difficult to infer the correctness of the decision on a single trial basis. Yet, the tails of the post-error synaptic distribution should allow in some cases to infer that an error has been made. It would be interesting to see in behavioral experiments whether the post-error effects can be related to the confidence in one's decision~\citep{Wei2015,Insabato2017}.

	\subsection*{Acknowledgments}
	
	We thank Jer\^{o}me Sackur, Jean-R\'emy Martin and Laurent Bonnasse-Gahot for useful discussions. We are grateful to the anonymous reviewers for helpful and constructive comments. 
	KB acknowledges a fellowship from the ENS Paris-Saclay.

				\bibliographystyle{apalike}

			\end{document}